\documentclass[
reprint, onlinecite, superscriptaddress,
amsmath,amssymb,
aps,prl
]{revtex4-1}

\usepackage{makecell}
\usepackage{graphicx}
\usepackage{dcolumn}
\usepackage{bm}
\usepackage{longtable}
\usepackage{physics}
\usepackage{amsmath}
\usepackage{multirow}

\usepackage{soul}
\usepackage[dvipsnames]{xcolor}

\begin{document}
\title{Crystal Symmetry Selected Pure Spin Photocurrent in Altermagnetic Insulators} 

\author{Ruizhi Dong}
\affiliation{School of Physics, Beijing Institute of Technology, Beijing, 100081, China}%
 
\author{Ranquan Cao}
\affiliation{School of Physics, Beijing Institute of Technology, Beijing, 100081, China}%
\affiliation{ Key Lab of Advanced Optoelectronic Quantum Architecture and Measurement (Ministry of Education), Beijing Institute of Technology, Beijing, 100081, China}%

\author{Dian Tan}
\affiliation{Shenzhen Institute for Quantum Science and Engineering and Department of Physics,
Southern University of Science and Technology, Shenzhen, 518055, China
}%

\author{Ruixiang Fei}
\email{rfei@bit.edu.cn}
\affiliation{School of Physics, Beijing Institute of Technology, Beijing, 100081, China}%
\affiliation{ Key Lab of Advanced Optoelectronic Quantum Architecture and Measurement (Ministry of Education), Beijing Institute of Technology, Beijing, 100081, China}%

\begin{abstract}

The generation of time-reversal-odd spin-current in metallic 
altermagnets has attracted considerable interest in spintronics.
However, producing pure spin-current in insulating materials
remains both challenging and desirable,
as insulating states are frequently found in antiferromagnets.
Nonlinear photogalvanic effects offer a promising method for generating spin-current in insulators.
We here
revealed that spin and 
charge photocurrents in altermagnets are protected by spin point
group symmetry. Unlike the photocurrents in parity-time
symmetric materials, where spin-orbit coupling (SOC) induces a 
significant charge current, the spin-current in altermagnets 
can exist as a pure spin current
along specific crystal directions regardless of 
SOC. We applied our predictions using
first-principles calculations to several distinct materials,
including wurtzite MnTe and multiferroic BiFeO$_3$. 
Additionally, we elucidated the previously overlooked 
linear-inject-current mechanism in BiFeO$_3$ induced by SOC, 
which may account for the enhanced bulk photovotaic effect
in multiferroics.
\end{abstract}

\maketitle

\textbf{\textit{Introduction}}-- 
Generating spin-polarized current, particularly pure spin current, in solid-state 
systems has received significant attention \cite{Baltz2018,Sinova2015}. 
A promising approach to generate pure spin-current is via the spin Hall effect
in metals \cite{Dyakonov1971,Murakami2003,Sinova2004} 
or the quantum spin Hall effect in topological insulators\cite{Kane2005,Bernevig2006}. 
These mechanisms typically do not require magnetic order but are
at the expense of relying on spin-orbit coupling (SOC). Recently, the altermagnetic
phase, characterized by alternating spin-splitting bands resulting from real-space 
rotation transformation, has been revealed \cite{Smejkal2020,Hayami2019,Yuan2020}.
Its significant spin splitting induced by strong non-relativistic exchange coupling,
along with its conservation of spin momentum are viewed as promising 
developments for spintronics \cite{Smejkal2020,Hayami2019,Yuan2020,Mazin2021,Bai2023} 
and possible applications \cite{Ouassou2023,Hariki2024,Leenders2024,Aoyama2024}.
For example, the broken time-reversal symmetry ($\mathcal{T}$) in 
altermagnets enables the generation of $\mathcal{T}$-odd spin 
current in metals\cite{Gonzalez-Hernandez2021,Ma2021} such as RuO$_2$, regardless of SOC. 
Generating spin current in insulators via such a DC
bias is impossible; however, spin-current in altermagnetic 
insulators is highly desirable, as insulating states 
constitute a large portion of antiferromagnets
\cite{Gao2023,Baltz2018}.

\begin{figure}[!]
\includegraphics[width=7.6cm]{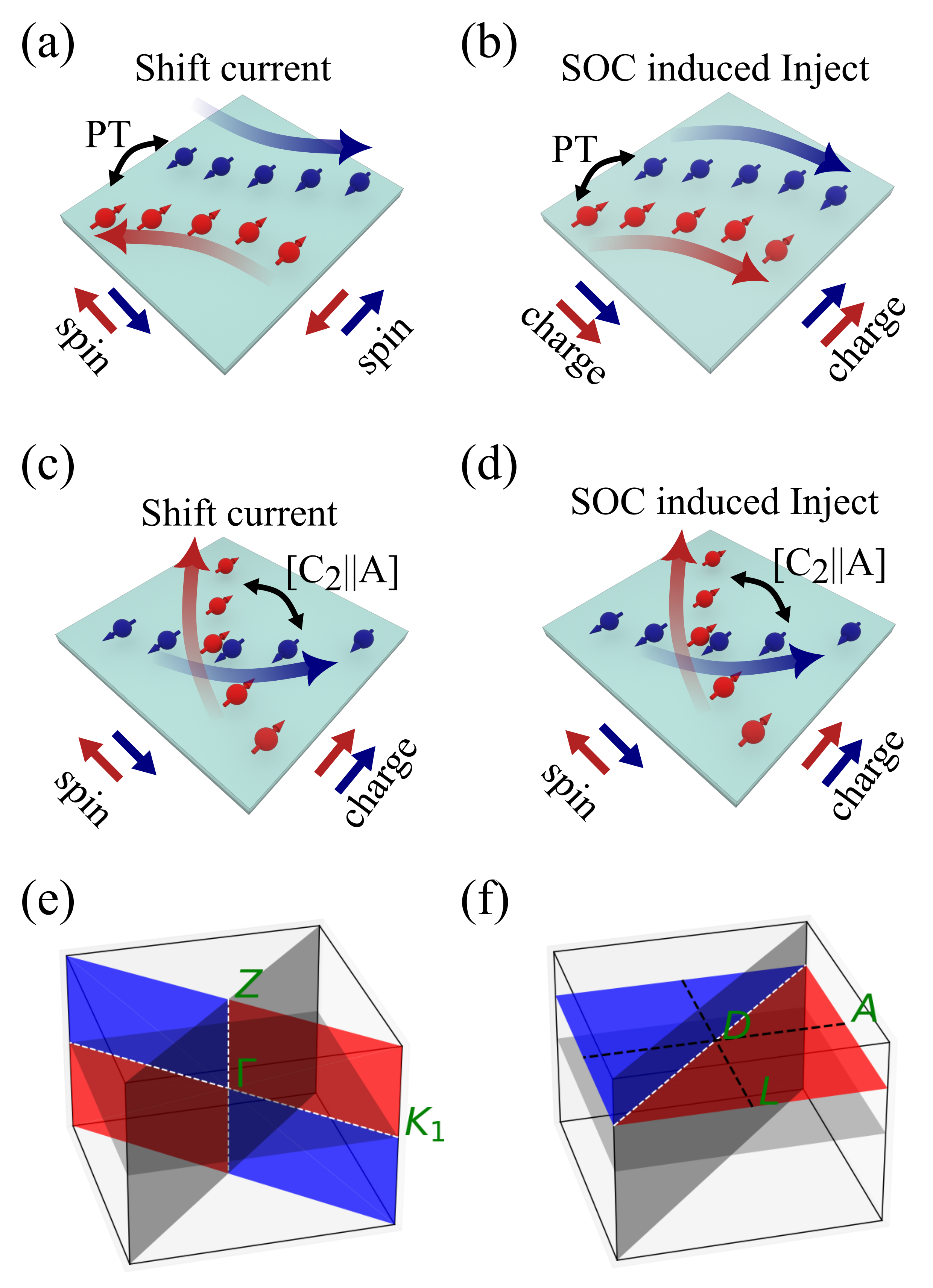}
\caption{ 
The schematic scenarios  illustrate the linearly 
polarized light driven spin-current (a) 
and the charge-current resulting from SOC (b) 
in $\mathcal{PT}$-AFMs, which disrupt the pure spin current. 
The crystal direction-dependent spin- or 
charge-current in altermagnet insulators for both 
shift-current (c) and inject-current (d).
The Brillouin zone of an altermagnet with a $^22$ 
spin point group, featuring two spin-splitting 
off-nodal planes $\Gamma ZK_1$ (e) 
and $DAL$ (f).
} \label{Figure1}
\end{figure}

Nonlinear photogalvanic effects are viewed as a promising approach
to generate spin-polarized current in noncentrosymmetric insulators\cite{Bhat2005,Ganichev2002,Zhou2007,Young2013,Fei2020,Xu2021,Fei2021,Song2021}. 
Notably, the two mechanisms, shift current and inject current\cite{sipe2000,VonBaltz1981}, 
exhibit different parity under time-reversal ($\mathcal{T}$) or parity-time ($\mathcal{PT}$) symmetry\cite{Fei2020,Xu2021,Ahn2020,Watanabe2021,Wang2019,Nagaosa2024}.
This allows them to produce spin or charge current, respectively, 
making them suitable for manipulation through optical linear or circular polarization.
For example, the linear-shift-current mechanism (referring to shift current induced by linearly polarized light)
would generate spin current in $\mathcal{PT}$-symmetric antiferromagnets ($\mathcal{PT}$-AFMs)\cite{Young2013}, 
while the circular-inject-current mechanism produces spin current\cite{Fei2021}. However,
both linear-inject-current and circular-shift-current mechanisms would generate 
charge current in $\mathcal{PT}$-AFMs with SOC\cite{Zhang2019,Fei2020a}, rendering pure spin current impossible. 
In contrast to $\mathcal{PT}$-AFM, where the spin-down and spin-up 
sublattices are connected by $\mathcal{PT}$ symmetry,
they are linked by real-space rotation transformation in altermagnets\cite{Smejkal2022}. 
This provides a solid platform for manipulating both spin and charge currents through its crystal symmetry using light.

In this work, we reveal the crystal symmetry-dependent
photo-driven spin and charge currents in altermagnets. 
Unlike $\mathcal{PT}$-AFM, 
the spin-up and spin-down sublattices in altermagnets 
are connected by real-space rotational transformation, 
allowing spin-up and spin-down electrons to exhibit covariant 
behavior in shift-current and inject-current mechanisms. 
Through symmetry analysis, we demonstrate that both mechanisms
can generate spin-current in specific directions, 
leading to pure spin-current regardless of SOC.
We validate these predictions through first-principles 
calculations of two experimentally accessible altmagnets:
wurtzite MnTe and multiferroic BiFeO$_3$. 
The spin currents, generated by these two mechanisms,
are significant and codirectional across a wide range of photonic energies. 

\textbf{\textit{Pure spin-current independent of SOC}}--  
In the case of coherent light illumination, the primary mechanisms
responsible for a second-order DC photocurrent
are the shift current $J_c=\sigma_{abc} E_a(\omega)E_b(-\omega) $
and inject current $J_c=\eta_{abc} E_a(\omega)E_b(-\omega)$ \cite{sipe2000,VonBaltz1981,Kraut1979}. 
Here, $E(\omega)$ represents the electrical field of incoming light with 
frequency $\omega$, with indices $a$ and $b$ 
denote the polarization directions of the light, and $c$ specifying the current direction. The shift current primarily characterizes the
polarization variance between the conduction and valence bands \cite{sipe2000,Ahn2020}:
\begin{align}
\begin{split}
\label{eq1}
\sigma_{a b c}=\frac{i \pi e^{3}}{\hbar^{2} \omega^{2}} \sum_{m n} \int d^{3}k &f_{m n} (v_{m n}^{a}(k) v_{n m ; c}^{b}(k)  \pm \\
& v_{m n}^{b}(k) v_{n m ; c}^{a}(k)) \delta\left(\omega-\omega_{m n}\right)  
\end{split}
\end{align}
where $f_{mn}$ denotes the difference in occupation numbers between two bands. 
$v_{mn}^{b}$ is the velocity matrix element, and
$v_{nm;c}^{b}=\frac{\partial v_{nm}^{b}}{\partial k^{c}}-i\left[A_{nn}^{c}-A_{mm}^{c}\right]v_{n m}^{b}\equiv v_{nm}^{b}R_{nm}^{c}$ 
defines the gauge-independent `generalized derivatives' of
velocity matrix element. $A_{nn}^{c}$ stands for the Berry connection 
of band $n$, while $R_{nm}^{c}$ denotes the shift vector.
Conversely, the inject current shows a linear dependence on the carrier lifetime ($\tau$)\cite{sipe2000}
\begin{align}
\begin{split}
\label{eq2}
\eta_{a b c}=\frac{-\pi e^{3}}{\hbar^{2} \omega^{2}} \sum_{mn} \int 
& d^{3}k 
   ((v_{m n}^{a}(k) v_{n m}^{b}(k) \pm v_{n m}^{a}(k) v_{m n}^{b}(k))\\
 & \cdot f_{mn}\left(v_{n n}^{c}-v_{m m}^{c}\right) \tau \delta\left(\omega-\omega_{m n}\right)
\end{split}
\end{align}

When examining the photo-driven spin-current, we substitute the velocity
matrix in the $c$ direction with 
$v_{m n}^{c}=\left\langle m\left|\left\{\sigma_{\alpha}, v_{c}\right\}\right| n\right\rangle$, where $\left\{\sigma_{\alpha}, v_{c}\right\}=\frac{1}{2}\left(v_{c} \sigma_{\alpha}+\sigma_{\alpha} v_{c}\right)$
with $\sigma_{\alpha}$ denoting the spin component along the $\alpha$-direction.
The ``+" in ``$\pm$" signifies linearly polarized light, 
while ``-" represents circularly polarized light for both shift current 
and inject current. This implies that the linearly polarized light and 
circularly polarized light cases depend on the real and imaginary parts 
of $v_{mn}^{a}(k)v_{nm}^{b}(k)$, respectively. The distinct parities 
of the real and imaginary parts of $v_{mn}^{a}(k) v_{nm}^{b}(k)$ under
various symmetry operators, such as $\mathcal{T}$ and $\mathcal{PT}$, 
are pivotal in determining the charge or spin current in photogalvanic effects.
Similar scenarios were found in the $\mathcal{T}$-odd 
or $\mathcal{T}$-even spin Hall effects
of antiferromagnetic metals\cite{Freimuth2014,Gonzalez-Hernandez2021}.

\begin{table*}[t!]
\renewcommand{\arraystretch}{1.1} 
\setlength{\tabcolsep}{12pt} 
\caption{
The spatial separation of spin and charge currents 
in altermagnets, contrasts with $\mathcal{PT}$-antiferromagnets, 
where charge current is always present due to SOC.
This is exemplified by scenarios where the spin-up 
and spin-down  sub-lattices in altermagnet are 
connected by rotational symmetry 
along the $z$-direction, e.g. $^22$ spin point group, 
with light polarization in the $xy$ plane. 
`Linearly' and `Circularly' refer to linearly 
and circularly polarized light, respectively.}
\label{tab:table1}
\centering
\begin{tabular}{cc|cc|cc} 
\hline
 & &
\multicolumn{2}{c|}{Linearly}& 
\multicolumn{2}{c}{Circularly}\\ 
\cline{3-6} 
 & & Shift & Inject & Shift & Inject \\ 
\hline
\multirow{2}{*}{$\mathcal{PT}$-AFM} & non-SOC  & spin & 0 & 0 & spin\tabularnewline
\cline{2-6} 
& SOC  & spin & charge & charge & spin\tabularnewline
\hline
\multirow{2}{*}{Altermagnet} & non-SOC  & \makecell{spin ($x$ or $y$) \\ charge($z$)} & 0 & 0 & \makecell{spin ($x$ or $y$) \\ charge($z$)}\tabularnewline
\cline{2-6} 
& SOC  & \makecell{spin ($x$ or $y$) \\ charge($z$)} & \makecell{spin ($x$ or $y$) \\ charge($z$)} & \makecell{spin ($x$ or $y$) \\ charge($z$)} & \makecell{spin ($x$ or $y$) \\ charge($z$)}\tabularnewline
\hline
\end{tabular}
\end{table*}


In antiferromagnets with $\mathcal{PT}$ symmetry,
the spin-up and spin-down electrons with energy degeneracy at the same 
$k$-point are connected by the $\mathcal{PT}$ symmetry in the reciprocal space. 
When applying the $\mathcal{PT}$ symmetry on spin $(s)$-dependent  
velocity matrix and shift vector, we obtained
$\mathcal{PT}\langle m,\uparrow,k|v|n,\uparrow, k\rangle=\langle n,\downarrow,k|v|m, \downarrow,k\rangle$ 
and $\mathcal{PT} R_{n m}^{c}(\uparrow, k)=-R_{n m}^{c}(\downarrow, k)$. 
As a result, 
the real and imaginary components of $v_{mn}^{a}(k)v_{nm}^{b}(k)$ 
exhibit $\mathcal{PT}$-even and $\mathcal{PT}$-odd characteristics 
for opposite spins, respectively. Following Eqs. 1 and 2, 
for linearly polarized light, which is associated 
with the $\mathcal{PT}$-even real component, the shift current 
and inject current generate spin-current and 
charge-current \cite{Young2013,Zhang2019,Fei2020a}, 
respectively  (See Table 1).
However, under circularly polarized light, this scenario reverses
entirely due to the $\mathcal{PT}$-odd characteristics of the imaginary component \cite{Fei2021}.

Note that $SU(2)$ spin-rotation symmetry $[\bar{C_2}||\mathcal{T}]$ is 
preserved in a system without SOC \cite{Smejkal2022,Liu2022}. Here the $\bar{C_2}$ is a $180^{\circ}$ 
rotation around an axis perpendicular to the spins, combined with
the spin-space inversion; the $\mathcal{T}$ is a time-reversal operator
in the real space which flips the sign of the crystal momentum. 
Applying the spin-rotation transformation
$[\bar{C_2}||\mathcal{T}]$  on spin (s)-dependent bands and velocity matrix, 
we obtained $[\bar{C_2}||\mathcal{T}]\epsilon_n(s,k)=\epsilon_n(s,-k)$ and 
$[\bar{C_2}||\mathcal{T}]\langle m,s,k|v|n,s,k\rangle=-\langle n,s,-k|v|m,s,-k\rangle$.
Consequently, the real-component of $v_{mn}^a (k)v_{nm}^b (k)$ and group velocity
are $[\bar{C_2}||\mathcal{T}]$-even and $[\bar{C_2}||\mathcal{T}]$-odd respectively, 
causing a zero charge-current from the linear-inject-current mechanism 
according to Eq.\ref{eq2}.
Similarly, the imaginary component of $v_{mn}^a (k)v_{nm}^b (k)$ and 
shift vector $R_{nm}^c (s,k)$ are $[\bar{C_2}||\mathcal{T}]$-odd 
and $[\bar{C_2}||\mathcal{T}]$-even respectively, causing a zero charge-current
from circular-shift-current mechanism according to Eq.\ref{eq1}. 
Thus, as depicted in Fig. 1(a), the linear-shift-current or the 
circular-inject-current would generate pure spin-current without any
charge-current when SOC is absent.  
 
The situation is entirely different for altermagnets,
as illustrated in Fig. 1b. The spin-up and spin-down electrons,
which are energy-degenerate at different $k$-point,
are connected by spin group transformation $[C_2||A]$ 
instead of the $\mathcal{PT}$ symmetry as in $\mathcal{PT}$-AFMs \cite{Smejkal2022}. 
Here, the transformation $C_2$ denotes a $180^{\circ}$ rotation 
in the spin-only space,
while transformation $A$ encompasses
solely real-space proper or improper rotations \cite{Smejkal2022}. 
In this scenario, the linear-shift-current mechanism can simultaneously 
generate spin and charge currents along different axes, as illustrated in Fig. 1(b).
This situation similarly applies to the circular-inject-current mechanism. 
If SOC is not considered, both the linear-inject-current and 
circular-shift-current result in zero charge or spin current, 
similar to the case in $\mathcal{PT}$-AFMs.


Without loss of generality, we discuss an altermagnet with 
a spin point group $^22$, where the spin-up and spin-down sublattices 
transform via $[C_2||C_{2z}]$. This transformation classifies
it as anisotropic $d$-wave altermagnets \cite{Smejkal2022}. 
As illustrated in Fig. 1c and 1d, the Brillouin zone is partitioned
into four regions by the gray-shaded nodal planes.
To demonstrate the influence of symmetry, 
we depict the spin polarization of one band, 
with spin-up and spin-down regions shown in red and blue, respectively.
For the spin-splitting on the off-nodal plane $\Gamma ZK_1$, 
the pattern exhibits inversion symmetry in $k$-space 
owing to the transformation $[\bar{C_2}||\mathcal{T}]$, which
persists in systems without SOC, regardless of crystal inversion. 
While on the off-nodal plane $DAL$, which is parallel to the spin-degenerate 
plane $\Gamma Y_1 S_1$ but with $k_z=\frac{\pi}{2c}$, the spin polarization 
demonstrates a two-fold rotation symmetry governed by 
spin-point group transformation $[C_2||C_{2z}]$.  

When applying the spin-rotation transformation $[\bar{C_2}||\mathcal{T}]$ on spin (s)-dependent
velocity matrix, we obtained
$[\bar{C_2}||\mathcal{T}]\langle m,s,k|v|n,s,k\rangle =-\langle n,s,-k|v|m,s,-k\rangle $. 
This results in zero charge-current governed by the linear-inject-current 
and circular-shift-current mechanisms, similar to that in $\mathcal{PT}$-AFMs. 
However, applying the transformation $[C_2||C_{2z}]$ on the 
velocity matrix results in distinct parities: even for 
the $x$- and $y$-directions, but odd for the $z$-direction. Specifically, 
we get: $[C_2||C_{2z}]\langle m,s,k|v_x|n,s,k\rangle =-\langle m,-s,(-k_x,-k_y,k_z)|v_x|n,-s,(-k_x,-k_y,k_z)\rangle $ 
and $[C_2||C_{2z}]\langle m,s,k|v_z|n,s,k\rangle =\langle m,-s,(-k_x,-k_y,k_z)|v_z|n,-s,(-k_x,-k_y,k_z)\rangle $. 
The shift vector $R_{nm}^c$ undergoes similar transformation relations
under $[C_2||C_{2z}]$. 
These properties suggest that spin-up and spin-down electrons, 
connected by $[C_2||C_{2z}]$, will travel in opposite directions 
along the $X$ or $Y$ axes but will move in the same direction
along the $Z$ axis. Consequently, the effective transportation of carriers in the
$Y$ or $X$ direction forms the spin-current, while the 
charge-current flows along the $Z$ direction.

When considering SOC, the spin point group $^22$ transitions 
to the magnetic point group $2'$ in collinear antiferromagnets, 
allowing for a non-zero linear-inject current (See Table 1). 
Under rotation symmetry $C_{2z}^{'}$, the spin-dependent velocity matrix
$\langle m,s,k|v_x|n,s,k\rangle$ is translated to
$-\langle m,-s,(-k_x,-k_y,k_z)|v_x|n,-s,(-k_x,-k_y,k_z)\rangle$,
resulting in $C_{2z}^{'}$-even real component of
$v_mn^a (k) v_nm^b (k)$ and $C_{2z}^{'}$-odd group velocity.
Consequently, a nonzero spin current is generated along the $X$ or $Y$ direction and a charge current along the $Z$ direction through the linear-inject-current mechanism. Thus, a pure spin current can be induced along the $X$ or $Y$ direction. This pattern holds for scenarios involving circularly polarized light with polarization in the $xy$ plane.


\textbf{\textit{Wurtzite MnTe}}-- 
The $\alpha$-MnTe with its NiAs-type structure possessing inversion
symmetry\cite{Lee2024,Krempasky2024,Osumi2024}, 
lacks the capability for generating second-order photo-driven
spin currents. 
In contrast, the wurtzite form of MnTe (spin point group $^26^2m^1m$), 
depicted in Fig. 2a, 
breaks inversion symmetry, leading to a notable photogalvanic effect.
Despite $\alpha$-MnTe being the energetically favorable structure \cite{Kriegner2017}, 
the wurtzite configuration remains metastable and has been successfully
synthesized and maintained at room temperature in experiments, demonstrating 
a moderate band-gap suitable for solar cell applications \cite{Siol2018,Mori2020}. 

\begin{figure}[!]
\includegraphics[width=7.8cm]{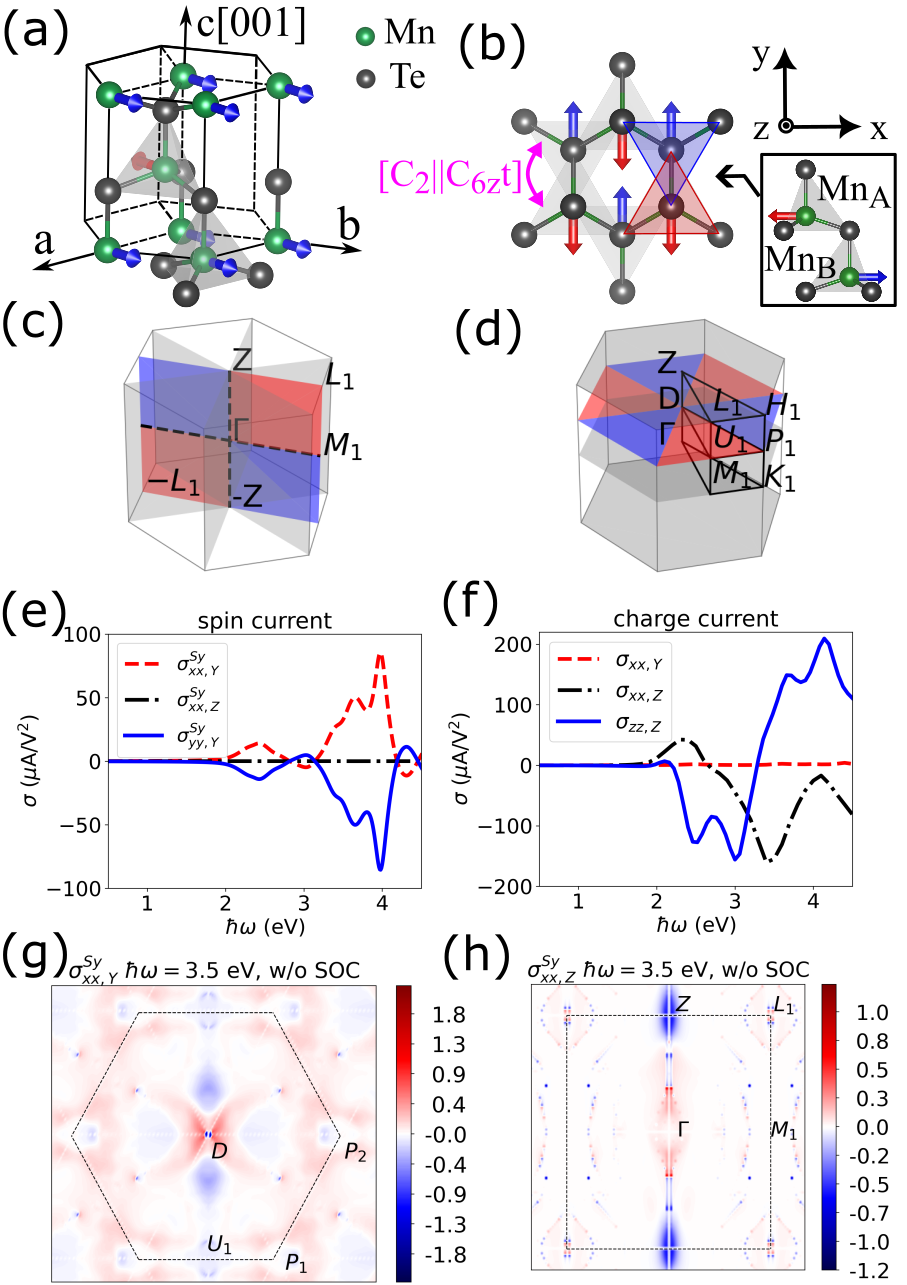}
\caption{
(a) The crystal structure of wurtzite MnTe, (b) the top view of 
two opposite-spin sublattices (Mn$_A$ and Mn$_B$), which are connected
through spin rotation combined with six-fold crystal rotation and
half-unit cell translation ($t$) along the $c$ axis. (c) the Brillouin zone
and the off-nodal plane $DU_1 P_1$ with $k_z=0.3\pi/c$,
(d) the $\Gamma ZM_1$ off-nodal plane. 
The spin-current (e) and charge-current (f) conductivities without SOC
originated from the linear-shift-current mechanism. 
The corresponding spin photoconductivity distribution 
on the $DU_1 P_1$ plane (g) and the $\Gamma ZM_1$ plane (h) 
without SOC for 3.5 eV photons.}
\label{Fig2}
\end{figure}

We discovered that the wurtzite structure exhibits characteristics
of a $g$-wave altermagnet, similar to the $\alpha$-phase. 
However, unlike the $\alpha$-phase, the two opposite-spin
sublattices (Mn$_A$ and Mn$_B$) in the wurtzite structure 
transform through $[C_2||C_{6z} t]$ (Fig. 2(b)), 
losing the $[C_2||M_z]$ symmetry that links the two
sublattices in the $\alpha$-phase(See supplement material \cite{supp,Kresse1996,Dudarev1998,Ihlefeld2008,Liechtenstein1995,Ederer2005,Lubk2009,perdew1996,Kiselev1963DetectionOM}).
In Figs.2c and 2d, the spin-degenerate nodal planes 
within the Brillouin zone are shaded in grey. 
The spin-degenerate plane $\Gamma K_1 M_1$ is protected 
by $[C_2||C_{6z}t]$ combined with $[\bar{C_2}||\mathcal{T}]$, 
and the plane $\Gamma K_1 Z$ is protected by $[C_2||C_{6z}t]$ 
combined with $[E||C_{3z}$]($E$ is the spin-space identity). 
The pattern on the off-nodal plane $\Gamma ZM_1$ exhibits
inversion symmetry 
owing to the spin-rotation operator $[\bar{C_2}||\mathcal{T}]$,
which persists in systems without SOC.
While for bands on the off-nodal plane $DU_1 P_1$, which is parallel
to $\Gamma K_1 M_1$ but with $k_z=\frac{0.3\pi}{c}$, 
the spin-splitting demonstrates a three-fold pattern governed 
by the crystal rotation operator $C_{3z}$, and the spin-up and spin-down
is connected by $[C_2||C_{6z}t]$, leading to the pattern shown in Fig. 2d. 

If SOC is not considered, our first-principles calculations 
confirm that the linear-inject-current and circular-shift-current 
mechanisms contribute neither to spin-current nor
charge-current. This is universally observed in 
$\mathcal{PT}$-AFMs and altermagnets, as listed in Table 1. 
However, as previously discussed, the linear-shift-current and 
circular-inject-current mechanisms can contribute to nonzero
spin or charge currents. These currents depend on the spin point group
$[C_2||C_{6z}t]$, which has a transformation $[C_2||C_{2z}t]$. 
To simplify the analysis, we apply the transformation $[C_2||C_{2z}t]$ 
to matrix $v_{mn}$. 
The resulting velocity matrices exhibit 
distinct parities: even for the x- and y-directions, but 
odd for the z-direction.
These characteristics indicate that the transport of carriers 
in the $Y$ or $X$ direction results in spin current, 
while the charge current flows along the $Z$ direction. 

The currents induced by the linear-shift-current mechanism
are illustrated in Fig. 2e and 2f, where the spin-current
$\sigma_{xx,Y}^{s_y}=-\sigma_{yy,Y}^{s_y}$ is nonzero, and the 
charge-current $\sigma_{xx,Z}$ is nonzero as well.
The $S_x$ and $S_z$ components spin-current are zero due to
the $y$-directional easy-axis of the Néel vector. Interestingly,
the spin and charge currents 
in the $X$-direction are zero due to mirror symmetry $[E||M_x]$.
Fig.2g and 2h illustrate the parities of shift-current conductivity 
under$[C_2||C_{2z}t]$ and $[\bar{C_2}||\mathcal{T}]$ operators, respectively. 
Both demonstrate an even behavior for the spin-conductivity
tensor $\sigma_{xx,Y}^{s_y}$, leading to a non-zero spin-current 
(See circular-inject-current in supplementary \cite{Fei2020}).
When SOC is considered, the inject-current and shift-current 
mechanisms still display the same behavior under the transformations
of spin point-group symmetry, resulting in pure spin currents
in a specific direction (see supplement information \cite{Fei2020}).
\begin{figure}
\includegraphics[width=8cm]{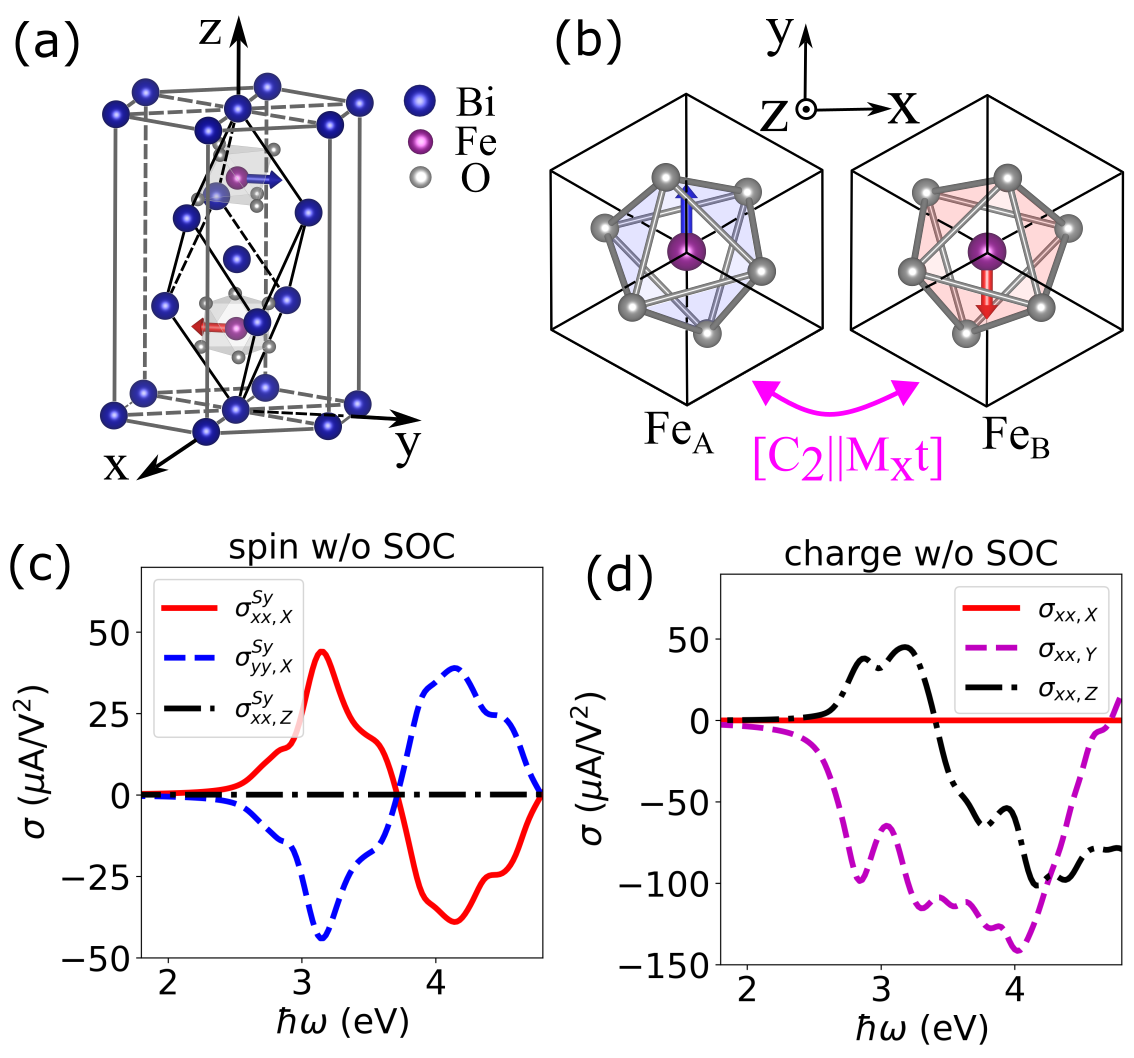} 
\caption{ 
(a) The crystal structure of multiferroic BiFeO$_3$
(b) the top view of two opposite-spin sublattices 
(Fe$_A$ and Fe$_B$), which are related by spin rotation
combined with mirror symmetry and half-unit cell translation ($t$). 
The spin-current (c) and charge-current (d) conductivities arise 
from the linear-shift-current mechanism without SOC.
The effective Hubbard U is set to 5 eV because the calculated imaginary permittivity best matches the experimental results for energies near the band gap.
} 
\label{Fig3}
\end{figure}

\textbf{\textit{Multiferroic BiFeO$_3$}}:
Bismuth ferrite (BFO) exhibits a ferroelectric
phase transition Curie temperature ($T_C$) of 1083–1103 $K$ 
and a G-type antiferromagnetic phase transition Néel
temperature ($T_N$) of 625–643 $K$, both significantly surpassing room
temperature\cite{Kiselev1963DetectionOM,Teague1970,Fischer1980a}.
This exceptional $T_C$ and $T_N$ make BFO
the leading candidate and widely recognized multiferroic material.
The polarization direction corresponds to the $[111]$ axis of
the rhombohedra primitive cell, aligning with the z-axis of the 
hexagonal structure, as illustrated in Fig. 3a. Additionally, 
the easy magnetization plane is perpendicular to the 
ferroelectric polarization axes. The two opposite-spin sublattices 
Fe$_A$ and Fe$_B$ are related by the transformation
$[C_2||M_x t]$ (Fig. 3b). Additionally, the symmetry
$[E||C_{3z}]$ places BFO in the spin-point group $^13^2m$,
classifying it as a g-wave altermagnet (see supplement information \cite{Fei2020}).

Our first-principles calculations confirm that only the 
linear-shift-current and circular-inject-current mechanisms 
contribute to the spin-current or charge-current in the absence of SOC.
Fig. 3c and 3d illustrate the spin-current 
and charge-current induced by the linear-shift-current mechanism,
respectively. For instance, when light is linearly polarized along
the $x$-direction,
spin-up and spin-down carriers effectively generate a $X$-direction spin-current,
while a $Y$ or $Z$ directions charge-current due to the 
transformation $[C_2||M_xt]$. 

\begin{figure}
\includegraphics[width=8cm]{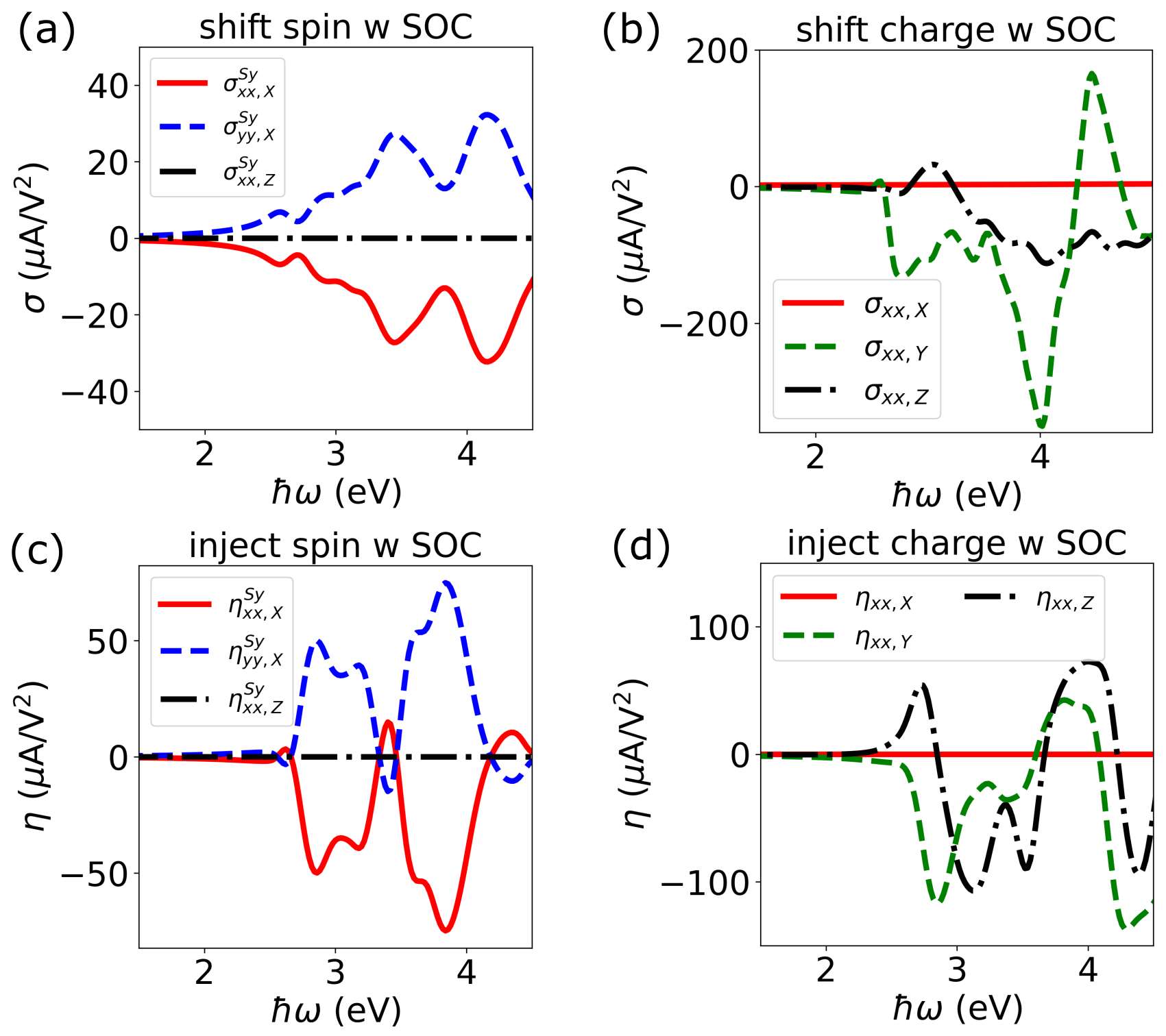} 
\caption{ The photo-driven spin- and charge-current of
multiferroic BiFeO$_3$ with SOC using $U_{eff}=5$ $eV$.
The spin-current (a) and charge-current (b) via the 
linear-shift-current mechanism.
And the spin-current (c) and charge-current (d) by the 
linear-inject-current mechanism.   
} 
\label{Fig4}
\end{figure}

In BFO, 
SOC plays a crucial role in 
light-induced spin currents and charge currents. 
Table 1 outlines the SOC enables linear-inject-current and
circular-shift-current mechanisms, as also confirmed in BFO through
first-principles calculations.
For example, 
Fig. 4a and 4b show the currents
from the linear-shift-current mechanism. 
In contrast to Figs. 3c and 3d, SOC can reverse 
the direction of the spin-current (e.g. $\sigma_{xxX}^{Sy}$)
by modifying the Berry connection of the bands. However,
whether conductivity components are nonzero or not 
remains unaffected, as they are determined by the
crystal symmetry.

Figs. 4c and 4d illustrate spin-current and
charge-current induced by linear-inject-current 
with SOC. Interestingly, the spin-current susceptibilities
induced by both shift-current and inject-current
are in the same direction across a
wide range of photonic energies, potentially resulting 
in a pure spin current, such as
$\sigma_{xxX}^{Sy}$ and $\sigma_{yyX}^{Sy}$. Interestingly, 
the charge current contributed by the inject-current 
is of the same order as that induced by the shift-current,
a factor overlooked in previous studies.
This additional contribution might explain the 
discrepancies between calculations
and experimental observations \cite{Young2-2012,Yang2010}.

In summary, we have demonstrated that the photo-driven 
spin and charge currents in altermagnets
directed separately due to the spin point group.
This behavior is fundamentally 
different from the $\mathcal{PT}$ symmetry in $\mathcal{PT}$-AFMs or $\mathcal{T}$-symmetry
in nonmagnetic materials, making altermagnets
uniquely capable of generating pure spin-current.
Furthermore, these crystal symmetry-selected pure spin-current
can be completely switched based on the type of light used,
whether linearly or circularly polarized, offering flexibility 
in tuning the spin current.
First-principles calculations further validate these 
findings in experimentally obtained materials, such as 
wurtzite MnTe and BiFeO$_3$, which could be beneficial 
for both spintronic devices and Néel vector detection. 

\begin{acknowledgments}
This work is supported by the National Natural Science Foundation of
China Grant No. 12204035.
\end{acknowledgments}

\bibliography{main}

\begin{thebibliography}{54}%
\makeatletter
\providecommand \@ifxundefined [1]{%
 \@ifx{#1\undefined}
}%
\providecommand \@ifnum [1]{%
 \ifnum #1\expandafter \@firstoftwo
 \else \expandafter \@secondoftwo
 \fi
}%
\providecommand \@ifx [1]{%
 \ifx #1\expandafter \@firstoftwo
 \else \expandafter \@secondoftwo
 \fi
}%
\providecommand \natexlab [1]{#1}%
\providecommand \enquote  [1]{``#1''}%
\providecommand \bibnamefont  [1]{#1}%
\providecommand \bibfnamefont [1]{#1}%
\providecommand \citenamefont [1]{#1}%
\providecommand \href@noop [0]{\@secondoftwo}%
\providecommand \href [0]{\begingroup \@sanitize@url \@href}%
\providecommand \@href[1]{\@@startlink{#1}\@@href}%
\providecommand \@@href[1]{\endgroup#1\@@endlink}%
\providecommand \@sanitize@url [0]{\catcode `\\12\catcode `\$12\catcode `\&12\catcode `\#12\catcode `\^12\catcode `\_12\catcode `\%12\relax}%
\providecommand \@@startlink[1]{}%
\providecommand \@@endlink[0]{}%
\providecommand \url  [0]{\begingroup\@sanitize@url \@url }%
\providecommand \@url [1]{\endgroup\@href {#1}{\urlprefix }}%
\providecommand \urlprefix  [0]{URL }%
\providecommand \Eprint [0]{\href }%
\providecommand \doibase [0]{http://dx.doi.org/}%
\providecommand \selectlanguage [0]{\@gobble}%
\providecommand \bibinfo  [0]{\@secondoftwo}%
\providecommand \bibfield  [0]{\@secondoftwo}%
\providecommand \translation [1]{[#1]}%
\providecommand \BibitemOpen [0]{}%
\providecommand \bibitemStop [0]{}%
\providecommand \bibitemNoStop [0]{.\EOS\space}%
\providecommand \EOS [0]{\spacefactor3000\relax}%
\providecommand \BibitemShut  [1]{\csname bibitem#1\endcsname}%
\let\auto@bib@innerbib\@empty
\bibitem [{\citenamefont {Baltz}\ \emph {et~al.}(2018)\citenamefont {Baltz}, \citenamefont {Manchon}, \citenamefont {Tsoi}, \citenamefont {Moriyama}, \citenamefont {Ono},\ and\ \citenamefont {Tserkovnyak}}]{Baltz2018}%
  \BibitemOpen
  \bibfield  {author} {\bibinfo {author} {\bibfnamefont {V.}~\bibnamefont {Baltz}}, \bibinfo {author} {\bibfnamefont {A.}~\bibnamefont {Manchon}}, \bibinfo {author} {\bibfnamefont {M.}~\bibnamefont {Tsoi}}, \bibinfo {author} {\bibfnamefont {T.}~\bibnamefont {Moriyama}}, \bibinfo {author} {\bibfnamefont {T.}~\bibnamefont {Ono}}, \ and\ \bibinfo {author} {\bibfnamefont {Y.}~\bibnamefont {Tserkovnyak}},\ }\href {\doibase 10.1103/RevModPhys.90.015005} {\bibfield  {journal} {\bibinfo  {journal} {Reviews of Modern Physics}\ }\textbf {\bibinfo {volume} {90}},\ \bibinfo {pages} {015005} (\bibinfo {year} {2018})}\BibitemShut {NoStop}%
\bibitem [{\citenamefont {Sinova}\ \emph {et~al.}(2015)\citenamefont {Sinova}, \citenamefont {Valenzuela}, \citenamefont {Wunderlich}, \citenamefont {Back},\ and\ \citenamefont {Jungwirth}}]{Sinova2015}%
  \BibitemOpen
  \bibfield  {author} {\bibinfo {author} {\bibfnamefont {J.}~\bibnamefont {Sinova}}, \bibinfo {author} {\bibfnamefont {S.~O.}\ \bibnamefont {Valenzuela}}, \bibinfo {author} {\bibfnamefont {J.}~\bibnamefont {Wunderlich}}, \bibinfo {author} {\bibfnamefont {C.}~\bibnamefont {Back}}, \ and\ \bibinfo {author} {\bibfnamefont {T.}~\bibnamefont {Jungwirth}},\ }\href {\doibase 10.1103/RevModPhys.87.1213} {\bibfield  {journal} {\bibinfo  {journal} {Reviews of Modern Physics}\ }\textbf {\bibinfo {volume} {87}},\ \bibinfo {pages} {1213} (\bibinfo {year} {2015})}\BibitemShut {NoStop}%
\bibitem [{\citenamefont {Dyakonov}\ and\ \citenamefont {Perel}(1971)}]{Dyakonov1971}%
  \BibitemOpen
  \bibfield  {author} {\bibinfo {author} {\bibfnamefont {M.}~\bibnamefont {Dyakonov}}\ and\ \bibinfo {author} {\bibfnamefont {V.}~\bibnamefont {Perel}},\ }\href {\doibase https://doi.org/10.1016/0375-9601(71)90196-4} {\bibfield  {journal} {\bibinfo  {journal} {Physics Letters A}\ }\textbf {\bibinfo {volume} {35}},\ \bibinfo {pages} {459} (\bibinfo {year} {1971})}\BibitemShut {NoStop}%
\bibitem [{\citenamefont {Murakami}\ \emph {et~al.}(2003)\citenamefont {Murakami}, \citenamefont {Nagaosa},\ and\ \citenamefont {Zhang}}]{Murakami2003}%
  \BibitemOpen
  \bibfield  {author} {\bibinfo {author} {\bibfnamefont {S.}~\bibnamefont {Murakami}}, \bibinfo {author} {\bibfnamefont {N.}~\bibnamefont {Nagaosa}}, \ and\ \bibinfo {author} {\bibfnamefont {S.-C.}\ \bibnamefont {Zhang}},\ }\href {\doibase 10.1126/science.1087128} {\bibfield  {journal} {\bibinfo  {journal} {Science}\ }\textbf {\bibinfo {volume} {301}},\ \bibinfo {pages} {1348} (\bibinfo {year} {2003})}\BibitemShut {NoStop}%
\bibitem [{\citenamefont {Sinova}\ \emph {et~al.}(2004)\citenamefont {Sinova}, \citenamefont {Culcer}, \citenamefont {Niu}, \citenamefont {Sinitsyn}, \citenamefont {Jungwirth},\ and\ \citenamefont {MacDonald}}]{Sinova2004}%
  \BibitemOpen
  \bibfield  {author} {\bibinfo {author} {\bibfnamefont {J.}~\bibnamefont {Sinova}}, \bibinfo {author} {\bibfnamefont {D.}~\bibnamefont {Culcer}}, \bibinfo {author} {\bibfnamefont {Q.}~\bibnamefont {Niu}}, \bibinfo {author} {\bibfnamefont {N.~A.}\ \bibnamefont {Sinitsyn}}, \bibinfo {author} {\bibfnamefont {T.}~\bibnamefont {Jungwirth}}, \ and\ \bibinfo {author} {\bibfnamefont {A.~H.}\ \bibnamefont {MacDonald}},\ }\href {\doibase 10.1103/PhysRevLett.92.126603} {\bibfield  {journal} {\bibinfo  {journal} {Physical Review Letters}\ }\textbf {\bibinfo {volume} {92}},\ \bibinfo {pages} {126603} (\bibinfo {year} {2004})}\BibitemShut {NoStop}%
\bibitem [{\citenamefont {Kane}\ and\ \citenamefont {Mele}(2005)}]{Kane2005}%
  \BibitemOpen
  \bibfield  {author} {\bibinfo {author} {\bibfnamefont {C.~L.}\ \bibnamefont {Kane}}\ and\ \bibinfo {author} {\bibfnamefont {E.~J.}\ \bibnamefont {Mele}},\ }\href {\doibase 10.1103/PhysRevLett.95.226801} {\bibfield  {journal} {\bibinfo  {journal} {Physical Review Letters}\ }\textbf {\bibinfo {volume} {95}},\ \bibinfo {pages} {226801} (\bibinfo {year} {2005})}\BibitemShut {NoStop}%
\bibitem [{\citenamefont {Bernevig}\ and\ \citenamefont {Zhang}(2006)}]{Bernevig2006}%
  \BibitemOpen
  \bibfield  {author} {\bibinfo {author} {\bibfnamefont {B.~A.}\ \bibnamefont {Bernevig}}\ and\ \bibinfo {author} {\bibfnamefont {S.-C.}\ \bibnamefont {Zhang}},\ }\href {\doibase 10.1103/PhysRevLett.96.106802} {\bibfield  {journal} {\bibinfo  {journal} {Physical Review Letters}\ }\textbf {\bibinfo {volume} {96}},\ \bibinfo {pages} {106802} (\bibinfo {year} {2006})}\BibitemShut {NoStop}%
\bibitem [{\citenamefont {Šmejkal}\ \emph {et~al.}(2020)\citenamefont {Šmejkal}, \citenamefont {González-Hernández}, \citenamefont {Jungwirth},\ and\ \citenamefont {Sinova}}]{Smejkal2020}%
  \BibitemOpen
  \bibfield  {author} {\bibinfo {author} {\bibfnamefont {L.}~\bibnamefont {Šmejkal}}, \bibinfo {author} {\bibfnamefont {R.}~\bibnamefont {González-Hernández}}, \bibinfo {author} {\bibfnamefont {T.}~\bibnamefont {Jungwirth}}, \ and\ \bibinfo {author} {\bibfnamefont {J.}~\bibnamefont {Sinova}},\ }\href {\doibase 10.1126/sciadv.aaz8809} {\bibfield  {journal} {\bibinfo  {journal} {Science Advances}\ }\textbf {\bibinfo {volume} {6}},\ \bibinfo {pages} {eaaz8809} (\bibinfo {year} {2020})}\BibitemShut {NoStop}%
\bibitem [{\citenamefont {Hayami}\ \emph {et~al.}(2019)\citenamefont {Hayami}, \citenamefont {Yanagi},\ and\ \citenamefont {Kusunose}}]{Hayami2019}%
  \BibitemOpen
  \bibfield  {author} {\bibinfo {author} {\bibfnamefont {S.}~\bibnamefont {Hayami}}, \bibinfo {author} {\bibfnamefont {Y.}~\bibnamefont {Yanagi}}, \ and\ \bibinfo {author} {\bibfnamefont {H.}~\bibnamefont {Kusunose}},\ }\href {\doibase 10.7566/JPSJ.88.123702} {\bibfield  {journal} {\bibinfo  {journal} {Journal of the Physical Society of Japan}\ }\textbf {\bibinfo {volume} {88}},\ \bibinfo {pages} {123702} (\bibinfo {year} {2019})}\BibitemShut {NoStop}%
\bibitem [{\citenamefont {Yuan}\ \emph {et~al.}(2020)\citenamefont {Yuan}, \citenamefont {Wang}, \citenamefont {Luo}, \citenamefont {Rashba},\ and\ \citenamefont {Zunger}}]{Yuan2020}%
  \BibitemOpen
  \bibfield  {author} {\bibinfo {author} {\bibfnamefont {L.-D.}\ \bibnamefont {Yuan}}, \bibinfo {author} {\bibfnamefont {Z.}~\bibnamefont {Wang}}, \bibinfo {author} {\bibfnamefont {J.-W.}\ \bibnamefont {Luo}}, \bibinfo {author} {\bibfnamefont {E.~I.}\ \bibnamefont {Rashba}}, \ and\ \bibinfo {author} {\bibfnamefont {A.}~\bibnamefont {Zunger}},\ }\href {\doibase 10.1103/PhysRevB.102.014422} {\bibfield  {journal} {\bibinfo  {journal} {Physical Review B}\ }\textbf {\bibinfo {volume} {102}},\ \bibinfo {pages} {014422} (\bibinfo {year} {2020})}\BibitemShut {NoStop}%
\bibitem [{\citenamefont {Mazin}\ \emph {et~al.}(2021)\citenamefont {Mazin}, \citenamefont {Koepernik}, \citenamefont {Johannes}, \citenamefont {González-Hernández},\ and\ \citenamefont {Šmejkal}}]{Mazin2021}%
  \BibitemOpen
  \bibfield  {author} {\bibinfo {author} {\bibfnamefont {I.~I.}\ \bibnamefont {Mazin}}, \bibinfo {author} {\bibfnamefont {K.}~\bibnamefont {Koepernik}}, \bibinfo {author} {\bibfnamefont {M.~D.}\ \bibnamefont {Johannes}}, \bibinfo {author} {\bibfnamefont {R.}~\bibnamefont {González-Hernández}}, \ and\ \bibinfo {author} {\bibfnamefont {L.}~\bibnamefont {Šmejkal}},\ }\href {\doibase 10.1073/pnas.2108924118} {\bibfield  {journal} {\bibinfo  {journal} {Proceedings of the National Academy of Sciences}\ }\textbf {\bibinfo {volume} {118}},\ \bibinfo {pages} {e2108924118} (\bibinfo {year} {2021})}\BibitemShut {NoStop}%
\bibitem [{\citenamefont {Bai}\ \emph {et~al.}(2023)\citenamefont {Bai}, \citenamefont {Zhang}, \citenamefont {Zhou}, \citenamefont {Chen}, \citenamefont {Wan}, \citenamefont {Han}, \citenamefont {Zhu}, \citenamefont {Liang}, \citenamefont {Su}, \citenamefont {Han}, \citenamefont {Pan},\ and\ \citenamefont {Song}}]{Bai2023}%
  \BibitemOpen
  \bibfield  {author} {\bibinfo {author} {\bibfnamefont {H.}~\bibnamefont {Bai}}, \bibinfo {author} {\bibfnamefont {Y.}~\bibnamefont {Zhang}}, \bibinfo {author} {\bibfnamefont {Y.}~\bibnamefont {Zhou}}, \bibinfo {author} {\bibfnamefont {P.}~\bibnamefont {Chen}}, \bibinfo {author} {\bibfnamefont {C.}~\bibnamefont {Wan}}, \bibinfo {author} {\bibfnamefont {L.}~\bibnamefont {Han}}, \bibinfo {author} {\bibfnamefont {W.}~\bibnamefont {Zhu}}, \bibinfo {author} {\bibfnamefont {S.}~\bibnamefont {Liang}}, \bibinfo {author} {\bibfnamefont {Y.}~\bibnamefont {Su}}, \bibinfo {author} {\bibfnamefont {X.}~\bibnamefont {Han}}, \bibinfo {author} {\bibfnamefont {F.}~\bibnamefont {Pan}}, \ and\ \bibinfo {author} {\bibfnamefont {C.}~\bibnamefont {Song}},\ }\href {\doibase 10.1103/PhysRevLett.130.216701} {\bibfield  {journal} {\bibinfo  {journal} {Physical Review Letters}\ }\textbf {\bibinfo {volume} {130}},\ \bibinfo {pages} {216701} (\bibinfo {year} {2023})}\BibitemShut {NoStop}%
\bibitem [{\citenamefont {Ouassou}\ \emph {et~al.}(2023)\citenamefont {Ouassou}, \citenamefont {Brataas},\ and\ \citenamefont {Linder}}]{Ouassou2023}%
  \BibitemOpen
  \bibfield  {author} {\bibinfo {author} {\bibfnamefont {J.~A.}\ \bibnamefont {Ouassou}}, \bibinfo {author} {\bibfnamefont {A.}~\bibnamefont {Brataas}}, \ and\ \bibinfo {author} {\bibfnamefont {J.}~\bibnamefont {Linder}},\ }\href {\doibase 10.1103/PhysRevLett.131.076003} {\bibfield  {journal} {\bibinfo  {journal} {Physical Review Letters}\ }\textbf {\bibinfo {volume} {131}},\ \bibinfo {pages} {076003} (\bibinfo {year} {2023})}\BibitemShut {NoStop}%
\bibitem [{\citenamefont {Hariki}\ \emph {et~al.}(2024)\citenamefont {Hariki}, \citenamefont {{Dal Din}}, \citenamefont {Amin}, \citenamefont {Yamaguchi}, \citenamefont {Badura}, \citenamefont {Kriegner}, \citenamefont {Edmonds}, \citenamefont {Campion}, \citenamefont {Wadley}, \citenamefont {Backes}, \citenamefont {Veiga}, \citenamefont {Dhesi}, \citenamefont {Springholz}, \citenamefont {{\v{S}}mejkal}, \citenamefont {V{\'{y}}born{\'{y}}}, \citenamefont {Jungwirth},\ and\ \citenamefont {Kune{\v{s}}}}]{Hariki2024}%
  \BibitemOpen
  \bibfield  {author} {\bibinfo {author} {\bibfnamefont {A.}~\bibnamefont {Hariki}}, \bibinfo {author} {\bibfnamefont {A.}~\bibnamefont {{Dal Din}}}, \bibinfo {author} {\bibfnamefont {O.~J.}\ \bibnamefont {Amin}}, \bibinfo {author} {\bibfnamefont {T.}~\bibnamefont {Yamaguchi}}, \bibinfo {author} {\bibfnamefont {A.}~\bibnamefont {Badura}}, \bibinfo {author} {\bibfnamefont {D.}~\bibnamefont {Kriegner}}, \bibinfo {author} {\bibfnamefont {K.~W.}\ \bibnamefont {Edmonds}}, \bibinfo {author} {\bibfnamefont {R.~P.}\ \bibnamefont {Campion}}, \bibinfo {author} {\bibfnamefont {P.}~\bibnamefont {Wadley}}, \bibinfo {author} {\bibfnamefont {D.}~\bibnamefont {Backes}}, \bibinfo {author} {\bibfnamefont {L.~S.~I.}\ \bibnamefont {Veiga}}, \bibinfo {author} {\bibfnamefont {S.~S.}\ \bibnamefont {Dhesi}}, \bibinfo {author} {\bibfnamefont {G.}~\bibnamefont {Springholz}}, \bibinfo {author} {\bibfnamefont {L.}~\bibnamefont {{\v{S}}mejkal}}, \bibinfo {author} {\bibfnamefont {K.}~\bibnamefont {V{\'{y}}born{\'{y}}}}, \bibinfo {author}
  {\bibfnamefont {T.}~\bibnamefont {Jungwirth}}, \ and\ \bibinfo {author} {\bibfnamefont {J.}~\bibnamefont {Kune{\v{s}}}},\ }\href {\doibase 10.1103/PhysRevLett.132.176701} {\bibfield  {journal} {\bibinfo  {journal} {Physical Review Letters}\ }\textbf {\bibinfo {volume} {132}},\ \bibinfo {pages} {176701} (\bibinfo {year} {2024})}\BibitemShut {NoStop}%
\bibitem [{\citenamefont {Leenders}\ \emph {et~al.}(2024)\citenamefont {Leenders}, \citenamefont {Afanasiev}, \citenamefont {Kimel},\ and\ \citenamefont {Mikhaylovskiy}}]{Leenders2024}%
  \BibitemOpen
  \bibfield  {author} {\bibinfo {author} {\bibfnamefont {R.~A.}\ \bibnamefont {Leenders}}, \bibinfo {author} {\bibfnamefont {D.}~\bibnamefont {Afanasiev}}, \bibinfo {author} {\bibfnamefont {A.~V.}\ \bibnamefont {Kimel}}, \ and\ \bibinfo {author} {\bibfnamefont {R.~V.}\ \bibnamefont {Mikhaylovskiy}},\ }\href {\doibase 10.1038/s41586-024-07448-3} {\bibfield  {journal} {\bibinfo  {journal} {Nature}\ }\textbf {\bibinfo {volume} {630}},\ \bibinfo {pages} {335} (\bibinfo {year} {2024})}\BibitemShut {NoStop}%
\bibitem [{\citenamefont {Aoyama}\ and\ \citenamefont {Ohgushi}(2024)}]{Aoyama2024}%
  \BibitemOpen
  \bibfield  {author} {\bibinfo {author} {\bibfnamefont {T.}~\bibnamefont {Aoyama}}\ and\ \bibinfo {author} {\bibfnamefont {K.}~\bibnamefont {Ohgushi}},\ }\href {\doibase 10.1103/PhysRevMaterials.8.L041402} {\bibfield  {journal} {\bibinfo  {journal} {Physical Review Materials}\ }\textbf {\bibinfo {volume} {8}},\ \bibinfo {pages} {L041402} (\bibinfo {year} {2024})}\BibitemShut {NoStop}%
\bibitem [{\citenamefont {Gonz{\'{a}}lez-Hern{\'{a}}ndez}\ \emph {et~al.}(2021)\citenamefont {Gonz{\'{a}}lez-Hern{\'{a}}ndez}, \citenamefont {{\v{S}}mejkal}, \citenamefont {V{\'{y}}born{\'{y}}}, \citenamefont {Yahagi}, \citenamefont {Sinova}, \citenamefont {Jungwirth},\ and\ \citenamefont {{\v{Z}}elezn{\'{y}}}}]{Gonzalez-Hernandez2021}%
  \BibitemOpen
  \bibfield  {author} {\bibinfo {author} {\bibfnamefont {R.}~\bibnamefont {Gonz{\'{a}}lez-Hern{\'{a}}ndez}}, \bibinfo {author} {\bibfnamefont {L.}~\bibnamefont {{\v{S}}mejkal}}, \bibinfo {author} {\bibfnamefont {K.}~\bibnamefont {V{\'{y}}born{\'{y}}}}, \bibinfo {author} {\bibfnamefont {Y.}~\bibnamefont {Yahagi}}, \bibinfo {author} {\bibfnamefont {J.}~\bibnamefont {Sinova}}, \bibinfo {author} {\bibfnamefont {T.}~\bibnamefont {Jungwirth}}, \ and\ \bibinfo {author} {\bibfnamefont {J.}~\bibnamefont {{\v{Z}}elezn{\'{y}}}},\ }\href {\doibase 10.1103/PhysRevLett.126.127701} {\bibfield  {journal} {\bibinfo  {journal} {Physical Review Letters}\ }\textbf {\bibinfo {volume} {126}},\ \bibinfo {pages} {127701} (\bibinfo {year} {2021})}\BibitemShut {NoStop}%
\bibitem [{\citenamefont {Ma}\ \emph {et~al.}(2021)\citenamefont {Ma}, \citenamefont {Hu}, \citenamefont {Li}, \citenamefont {Liu}, \citenamefont {Yao}, \citenamefont {Jia},\ and\ \citenamefont {Liu}}]{Ma2021}%
  \BibitemOpen
  \bibfield  {author} {\bibinfo {author} {\bibfnamefont {H.-Y.}\ \bibnamefont {Ma}}, \bibinfo {author} {\bibfnamefont {M.}~\bibnamefont {Hu}}, \bibinfo {author} {\bibfnamefont {N.}~\bibnamefont {Li}}, \bibinfo {author} {\bibfnamefont {J.}~\bibnamefont {Liu}}, \bibinfo {author} {\bibfnamefont {W.}~\bibnamefont {Yao}}, \bibinfo {author} {\bibfnamefont {J.-F.}\ \bibnamefont {Jia}}, \ and\ \bibinfo {author} {\bibfnamefont {J.}~\bibnamefont {Liu}},\ }\href {\doibase 10.1038/s41467-021-23127-7} {\bibfield  {journal} {\bibinfo  {journal} {Nature Communications}\ }\textbf {\bibinfo {volume} {12}},\ \bibinfo {pages} {2846} (\bibinfo {year} {2021})}\BibitemShut {NoStop}%
\bibitem [{\citenamefont {Gao}\ \emph {et~al.}(2023)\citenamefont {Gao}, \citenamefont {Qu}, \citenamefont {Zeng}, \citenamefont {Liu}, \citenamefont {Wen}, \citenamefont {Sun}, \citenamefont {Guo},\ and\ \citenamefont {Lu}}]{Gao2023}%
  \BibitemOpen
  \bibfield  {author} {\bibinfo {author} {\bibfnamefont {Z.-F.}\ \bibnamefont {Gao}}, \bibinfo {author} {\bibfnamefont {S.}~\bibnamefont {Qu}}, \bibinfo {author} {\bibfnamefont {B.}~\bibnamefont {Zeng}}, \bibinfo {author} {\bibfnamefont {Y.}~\bibnamefont {Liu}}, \bibinfo {author} {\bibfnamefont {J.-R.}\ \bibnamefont {Wen}}, \bibinfo {author} {\bibfnamefont {H.}~\bibnamefont {Sun}}, \bibinfo {author} {\bibfnamefont {P.-J.}\ \bibnamefont {Guo}}, \ and\ \bibinfo {author} {\bibfnamefont {Z.-Y.}\ \bibnamefont {Lu}},\ }\href {http://arxiv.org/abs/2311.04418} {\bibfield  {journal} {\bibinfo  {journal} {arXiv}\ ,\ \bibinfo {pages} {2311.04418}} (\bibinfo {year} {2023})}\BibitemShut {NoStop}%
\bibitem [{\citenamefont {Bhat}\ \emph {et~al.}(2005)\citenamefont {Bhat}, \citenamefont {Nastos}, \citenamefont {Najmaie},\ and\ \citenamefont {Sipe}}]{Bhat2005}%
  \BibitemOpen
  \bibfield  {author} {\bibinfo {author} {\bibfnamefont {R.~D.~R.}\ \bibnamefont {Bhat}}, \bibinfo {author} {\bibfnamefont {F.}~\bibnamefont {Nastos}}, \bibinfo {author} {\bibfnamefont {A.}~\bibnamefont {Najmaie}}, \ and\ \bibinfo {author} {\bibfnamefont {J.~E.}\ \bibnamefont {Sipe}},\ }\href {\doibase 10.1103/PhysRevLett.94.096603} {\bibfield  {journal} {\bibinfo  {journal} {Physical Review Letters}\ }\textbf {\bibinfo {volume} {94}},\ \bibinfo {pages} {096603} (\bibinfo {year} {2005})}\BibitemShut {NoStop}%
\bibitem [{\citenamefont {Ganichev}\ \emph {et~al.}(2002)\citenamefont {Ganichev}, \citenamefont {Ivchenko}, \citenamefont {Bel'kov}, \citenamefont {Tarasenko}, \citenamefont {Sollinger}, \citenamefont {Weiss}, \citenamefont {Wegscheider},\ and\ \citenamefont {Prettl}}]{Ganichev2002}%
  \BibitemOpen
  \bibfield  {author} {\bibinfo {author} {\bibfnamefont {S.~D.}\ \bibnamefont {Ganichev}}, \bibinfo {author} {\bibfnamefont {E.~L.}\ \bibnamefont {Ivchenko}}, \bibinfo {author} {\bibfnamefont {V.~V.}\ \bibnamefont {Bel'kov}}, \bibinfo {author} {\bibfnamefont {S.~A.}\ \bibnamefont {Tarasenko}}, \bibinfo {author} {\bibfnamefont {M.}~\bibnamefont {Sollinger}}, \bibinfo {author} {\bibfnamefont {D.}~\bibnamefont {Weiss}}, \bibinfo {author} {\bibfnamefont {W.}~\bibnamefont {Wegscheider}}, \ and\ \bibinfo {author} {\bibfnamefont {W.}~\bibnamefont {Prettl}},\ }\href {\doibase 10.1038/417153a} {\bibfield  {journal} {\bibinfo  {journal} {Nature}\ }\textbf {\bibinfo {volume} {417}},\ \bibinfo {pages} {153} (\bibinfo {year} {2002})}\BibitemShut {NoStop}%
\bibitem [{\citenamefont {Zhou}\ and\ \citenamefont {Shen}(2007)}]{Zhou2007}%
  \BibitemOpen
  \bibfield  {author} {\bibinfo {author} {\bibfnamefont {B.}~\bibnamefont {Zhou}}\ and\ \bibinfo {author} {\bibfnamefont {S.-Q.}\ \bibnamefont {Shen}},\ }\href {\doibase 10.1103/PhysRevB.75.045339} {\bibfield  {journal} {\bibinfo  {journal} {Physical Review B}\ }\textbf {\bibinfo {volume} {75}},\ \bibinfo {pages} {045339} (\bibinfo {year} {2007})}\BibitemShut {NoStop}%
\bibitem [{\citenamefont {Young}\ \emph {et~al.}(2013)\citenamefont {Young}, \citenamefont {Zheng},\ and\ \citenamefont {Rappe}}]{Young2013}%
  \BibitemOpen
  \bibfield  {author} {\bibinfo {author} {\bibfnamefont {S.~M.}\ \bibnamefont {Young}}, \bibinfo {author} {\bibfnamefont {F.}~\bibnamefont {Zheng}}, \ and\ \bibinfo {author} {\bibfnamefont {A.~M.}\ \bibnamefont {Rappe}},\ }\href {\doibase 10.1103/PhysRevLett.110.057201} {\bibfield  {journal} {\bibinfo  {journal} {Physical Review Letters}\ }\textbf {\bibinfo {volume} {110}},\ \bibinfo {pages} {057201} (\bibinfo {year} {2013})}\BibitemShut {NoStop}%
\bibitem [{\citenamefont {Fei}\ \emph {et~al.}(2020{\natexlab{a}})\citenamefont {Fei}, \citenamefont {Lu},\ and\ \citenamefont {Yang}}]{Fei2020}%
  \BibitemOpen
  \bibfield  {author} {\bibinfo {author} {\bibfnamefont {R.}~\bibnamefont {Fei}}, \bibinfo {author} {\bibfnamefont {X.}~\bibnamefont {Lu}}, \ and\ \bibinfo {author} {\bibfnamefont {L.}~\bibnamefont {Yang}},\ }\href {http://arxiv.org/abs/2006.10690} {\bibfield  {journal} {\bibinfo  {journal} {arXiv}\ ,\ \bibinfo {pages} {2006.10690}} (\bibinfo {year} {2020}{\natexlab{a}})}\BibitemShut {NoStop}%
\bibitem [{\citenamefont {Xu}\ \emph {et~al.}(2021)\citenamefont {Xu}, \citenamefont {Wang}, \citenamefont {Zhou},\ and\ \citenamefont {Li}}]{Xu2021}%
  \BibitemOpen
  \bibfield  {author} {\bibinfo {author} {\bibfnamefont {H.}~\bibnamefont {Xu}}, \bibinfo {author} {\bibfnamefont {H.}~\bibnamefont {Wang}}, \bibinfo {author} {\bibfnamefont {J.}~\bibnamefont {Zhou}}, \ and\ \bibinfo {author} {\bibfnamefont {J.}~\bibnamefont {Li}},\ }\href {\doibase 10.1038/s41467-021-24541-7} {\bibfield  {journal} {\bibinfo  {journal} {Nature Communications}\ }\textbf {\bibinfo {volume} {12}},\ \bibinfo {pages} {4330} (\bibinfo {year} {2021})}\BibitemShut {NoStop}%
\bibitem [{\citenamefont {Fei}\ \emph {et~al.}(2021)\citenamefont {Fei}, \citenamefont {Song}, \citenamefont {Pusey-Nazzaro},\ and\ \citenamefont {Yang}}]{Fei2021}%
  \BibitemOpen
  \bibfield  {author} {\bibinfo {author} {\bibfnamefont {R.}~\bibnamefont {Fei}}, \bibinfo {author} {\bibfnamefont {W.}~\bibnamefont {Song}}, \bibinfo {author} {\bibfnamefont {L.}~\bibnamefont {Pusey-Nazzaro}}, \ and\ \bibinfo {author} {\bibfnamefont {L.}~\bibnamefont {Yang}},\ }\href {\doibase 10.1103/PhysRevLett.127.207402} {\bibfield  {journal} {\bibinfo  {journal} {Physical Review Letters}\ }\textbf {\bibinfo {volume} {127}},\ \bibinfo {pages} {207402} (\bibinfo {year} {2021})}\BibitemShut {NoStop}%
\bibitem [{\citenamefont {Song}\ \emph {et~al.}(2021)\citenamefont {Song}, \citenamefont {Anderson}, \citenamefont {Tu}, \citenamefont {Seyler}, \citenamefont {Taniguchi}, \citenamefont {Watanabe}, \citenamefont {McGuire}, \citenamefont {Li}, \citenamefont {Cao}, \citenamefont {Xiao}, \citenamefont {Yao},\ and\ \citenamefont {Xu}}]{Song2021}%
  \BibitemOpen
  \bibfield  {author} {\bibinfo {author} {\bibfnamefont {T.}~\bibnamefont {Song}}, \bibinfo {author} {\bibfnamefont {E.}~\bibnamefont {Anderson}}, \bibinfo {author} {\bibfnamefont {M.~W.-Y.}\ \bibnamefont {Tu}}, \bibinfo {author} {\bibfnamefont {K.}~\bibnamefont {Seyler}}, \bibinfo {author} {\bibfnamefont {T.}~\bibnamefont {Taniguchi}}, \bibinfo {author} {\bibfnamefont {K.}~\bibnamefont {Watanabe}}, \bibinfo {author} {\bibfnamefont {M.~A.}\ \bibnamefont {McGuire}}, \bibinfo {author} {\bibfnamefont {X.}~\bibnamefont {Li}}, \bibinfo {author} {\bibfnamefont {T.}~\bibnamefont {Cao}}, \bibinfo {author} {\bibfnamefont {D.}~\bibnamefont {Xiao}}, \bibinfo {author} {\bibfnamefont {W.}~\bibnamefont {Yao}}, \ and\ \bibinfo {author} {\bibfnamefont {X.}~\bibnamefont {Xu}},\ }\href {\doibase 10.1126/sciadv.abg8094} {\bibfield  {journal} {\bibinfo  {journal} {Science Advances}\ }\textbf {\bibinfo {volume} {7}},\ \bibinfo {pages} {eabg8094} (\bibinfo {year} {2021})}\BibitemShut {NoStop}%
\bibitem [{\citenamefont {Sipe}\ and\ \citenamefont {Shkrebtii}(2000)}]{sipe2000}%
  \BibitemOpen
  \bibfield  {author} {\bibinfo {author} {\bibfnamefont {J.~E.}\ \bibnamefont {Sipe}}\ and\ \bibinfo {author} {\bibfnamefont {A.~I.}\ \bibnamefont {Shkrebtii}},\ }\href {\doibase 10.1103/PhysRevB.61.5337} {\bibfield  {journal} {\bibinfo  {journal} {Physical Review B}\ }\textbf {\bibinfo {volume} {61}},\ \bibinfo {pages} {5337} (\bibinfo {year} {2000})}\BibitemShut {NoStop}%
\bibitem [{\citenamefont {von Baltz}\ and\ \citenamefont {Kraut}(1981)}]{VonBaltz1981}%
  \BibitemOpen
  \bibfield  {author} {\bibinfo {author} {\bibfnamefont {R.}~\bibnamefont {von Baltz}}\ and\ \bibinfo {author} {\bibfnamefont {W.}~\bibnamefont {Kraut}},\ }\href {\doibase 10.1103/PhysRevB.23.5590} {\bibfield  {journal} {\bibinfo  {journal} {Physical Review B}\ }\textbf {\bibinfo {volume} {23}},\ \bibinfo {pages} {5590} (\bibinfo {year} {1981})}\BibitemShut {NoStop}%
\bibitem [{\citenamefont {Ahn}\ \emph {et~al.}(2020)\citenamefont {Ahn}, \citenamefont {Guo},\ and\ \citenamefont {Nagaosa}}]{Ahn2020}%
  \BibitemOpen
  \bibfield  {author} {\bibinfo {author} {\bibfnamefont {J.}~\bibnamefont {Ahn}}, \bibinfo {author} {\bibfnamefont {G.-Y.}\ \bibnamefont {Guo}}, \ and\ \bibinfo {author} {\bibfnamefont {N.}~\bibnamefont {Nagaosa}},\ }\href {\doibase 10.1103/PhysRevX.10.041041} {\bibfield  {journal} {\bibinfo  {journal} {Physical Review X}\ }\textbf {\bibinfo {volume} {10}},\ \bibinfo {pages} {041041} (\bibinfo {year} {2020})}\BibitemShut {NoStop}%
\bibitem [{\citenamefont {Watanabe}\ and\ \citenamefont {Yanase}(2021)}]{Watanabe2021}%
  \BibitemOpen
  \bibfield  {author} {\bibinfo {author} {\bibfnamefont {H.}~\bibnamefont {Watanabe}}\ and\ \bibinfo {author} {\bibfnamefont {Y.}~\bibnamefont {Yanase}},\ }\href {\doibase 10.1103/PhysRevX.11.011001} {\bibfield  {journal} {\bibinfo  {journal} {Physical Review X}\ }\textbf {\bibinfo {volume} {11}},\ \bibinfo {pages} {011001} (\bibinfo {year} {2021})}\BibitemShut {NoStop}%
\bibitem [{\citenamefont {Wang}\ and\ \citenamefont {Qian}(2019)}]{Wang2019}%
  \BibitemOpen
  \bibfield  {author} {\bibinfo {author} {\bibfnamefont {H.}~\bibnamefont {Wang}}\ and\ \bibinfo {author} {\bibfnamefont {X.}~\bibnamefont {Qian}},\ }\href {\doibase 10.1126/sciadv.aav9743} {\bibfield  {journal} {\bibinfo  {journal} {Science Advances}\ }\textbf {\bibinfo {volume} {5}},\ \bibinfo {pages} {eaav9743} (\bibinfo {year} {2019})}\BibitemShut {NoStop}%
\bibitem [{\citenamefont {Nagaosa}\ and\ \citenamefont {Yanase}(2024)}]{Nagaosa2024}%
  \BibitemOpen
  \bibfield  {author} {\bibinfo {author} {\bibfnamefont {N.}~\bibnamefont {Nagaosa}}\ and\ \bibinfo {author} {\bibfnamefont {Y.}~\bibnamefont {Yanase}},\ }\href {\doibase 10.1146/annurev-conmatphys-032822-033734} {\bibfield  {journal} {\bibinfo  {journal} {Annual Review of Condensed Matter Physics}\ }\textbf {\bibinfo {volume} {15}},\ \bibinfo {pages} {63} (\bibinfo {year} {2024})}\BibitemShut {NoStop}%
\bibitem [{\citenamefont {Zhang}\ \emph {et~al.}(2019)\citenamefont {Zhang}, \citenamefont {Holder}, \citenamefont {Ishizuka}, \citenamefont {de~Juan}, \citenamefont {Nagaosa}, \citenamefont {Felser},\ and\ \citenamefont {Yan}}]{Zhang2019}%
  \BibitemOpen
  \bibfield  {author} {\bibinfo {author} {\bibfnamefont {Y.}~\bibnamefont {Zhang}}, \bibinfo {author} {\bibfnamefont {T.}~\bibnamefont {Holder}}, \bibinfo {author} {\bibfnamefont {H.}~\bibnamefont {Ishizuka}}, \bibinfo {author} {\bibfnamefont {F.}~\bibnamefont {de~Juan}}, \bibinfo {author} {\bibfnamefont {N.}~\bibnamefont {Nagaosa}}, \bibinfo {author} {\bibfnamefont {C.}~\bibnamefont {Felser}}, \ and\ \bibinfo {author} {\bibfnamefont {B.}~\bibnamefont {Yan}},\ }\href {\doibase 10.1038/s41467-019-11832-3} {\bibfield  {journal} {\bibinfo  {journal} {Nature Communications}\ }\textbf {\bibinfo {volume} {10}},\ \bibinfo {pages} {3783} (\bibinfo {year} {2019})}\BibitemShut {NoStop}%
\bibitem [{\citenamefont {Fei}\ \emph {et~al.}(2020{\natexlab{b}})\citenamefont {Fei}, \citenamefont {Song},\ and\ \citenamefont {Yang}}]{Fei2020a}%
  \BibitemOpen
  \bibfield  {author} {\bibinfo {author} {\bibfnamefont {R.}~\bibnamefont {Fei}}, \bibinfo {author} {\bibfnamefont {W.}~\bibnamefont {Song}}, \ and\ \bibinfo {author} {\bibfnamefont {L.}~\bibnamefont {Yang}},\ }\href {\doibase 10.1103/PhysRevB.102.035440} {\bibfield  {journal} {\bibinfo  {journal} {Physical Review B}\ }\textbf {\bibinfo {volume} {102}},\ \bibinfo {pages} {035440} (\bibinfo {year} {2020}{\natexlab{b}})}\BibitemShut {NoStop}%
\bibitem [{\citenamefont {{\v{S}}mejkal}\ \emph {et~al.}(2022)\citenamefont {{\v{S}}mejkal}, \citenamefont {Sinova},\ and\ \citenamefont {Jungwirth}}]{Smejkal2022}%
  \BibitemOpen
  \bibfield  {author} {\bibinfo {author} {\bibfnamefont {L.}~\bibnamefont {{\v{S}}mejkal}}, \bibinfo {author} {\bibfnamefont {J.}~\bibnamefont {Sinova}}, \ and\ \bibinfo {author} {\bibfnamefont {T.}~\bibnamefont {Jungwirth}},\ }\href {\doibase 10.1103/PhysRevX.12.031042} {\bibfield  {journal} {\bibinfo  {journal} {Physical Review X}\ }\textbf {\bibinfo {volume} {12}},\ \bibinfo {pages} {031042} (\bibinfo {year} {2022})}\BibitemShut {NoStop}%
\bibitem [{\citenamefont {Kraut}\ and\ \citenamefont {Baltz}(1979)}]{Kraut1979}%
  \BibitemOpen
  \bibfield  {author} {\bibinfo {author} {\bibfnamefont {W.}~\bibnamefont {Kraut}}\ and\ \bibinfo {author} {\bibfnamefont {R.~V.}\ \bibnamefont {Baltz}},\ }\href@noop {} {\bibfield  {journal} {\bibinfo  {journal} {Physical Review B}\ }\textbf {\bibinfo {volume} {19}},\ \bibinfo {pages} {1548} (\bibinfo {year} {1979})}\BibitemShut {NoStop}%
\bibitem [{\citenamefont {Freimuth}\ \emph {et~al.}(2014)\citenamefont {Freimuth}, \citenamefont {Bl\"ugel},\ and\ \citenamefont {Mokrousov}}]{Freimuth2014}%
  \BibitemOpen
  \bibfield  {author} {\bibinfo {author} {\bibfnamefont {F.}~\bibnamefont {Freimuth}}, \bibinfo {author} {\bibfnamefont {S.}~\bibnamefont {Bl\"ugel}}, \ and\ \bibinfo {author} {\bibfnamefont {Y.}~\bibnamefont {Mokrousov}},\ }\href {\doibase 10.1103/PhysRevB.90.174423} {\bibfield  {journal} {\bibinfo  {journal} {Phys. Rev. B}\ }\textbf {\bibinfo {volume} {90}},\ \bibinfo {pages} {174423} (\bibinfo {year} {2014})}\BibitemShut {NoStop}%
\bibitem [{\citenamefont {Liu}\ \emph {et~al.}(2022)\citenamefont {Liu}, \citenamefont {Li}, \citenamefont {Han}, \citenamefont {Wan},\ and\ \citenamefont {Liu}}]{Liu2022}%
  \BibitemOpen
  \bibfield  {author} {\bibinfo {author} {\bibfnamefont {P.}~\bibnamefont {Liu}}, \bibinfo {author} {\bibfnamefont {J.}~\bibnamefont {Li}}, \bibinfo {author} {\bibfnamefont {J.}~\bibnamefont {Han}}, \bibinfo {author} {\bibfnamefont {X.}~\bibnamefont {Wan}}, \ and\ \bibinfo {author} {\bibfnamefont {Q.}~\bibnamefont {Liu}},\ }\href {\doibase 10.1103/PhysRevX.12.021016} {\bibfield  {journal} {\bibinfo  {journal} {Physical Review X}\ }\textbf {\bibinfo {volume} {12}},\ \bibinfo {pages} {021016} (\bibinfo {year} {2022})}\BibitemShut {NoStop}%
\bibitem [{\citenamefont {Lee}\ \emph {et~al.}(2024)\citenamefont {Lee}, \citenamefont {Lee}, \citenamefont {Jung}, \citenamefont {Jung}, \citenamefont {Kim}, \citenamefont {Lee}, \citenamefont {Seok}, \citenamefont {Kim}, \citenamefont {Park}, \citenamefont {{\v{S}}mejkal}, \citenamefont {Kang},\ and\ \citenamefont {Kim}}]{Lee2024}%
  \BibitemOpen
  \bibfield  {author} {\bibinfo {author} {\bibfnamefont {S.}~\bibnamefont {Lee}}, \bibinfo {author} {\bibfnamefont {S.}~\bibnamefont {Lee}}, \bibinfo {author} {\bibfnamefont {S.}~\bibnamefont {Jung}}, \bibinfo {author} {\bibfnamefont {J.}~\bibnamefont {Jung}}, \bibinfo {author} {\bibfnamefont {D.}~\bibnamefont {Kim}}, \bibinfo {author} {\bibfnamefont {Y.}~\bibnamefont {Lee}}, \bibinfo {author} {\bibfnamefont {B.}~\bibnamefont {Seok}}, \bibinfo {author} {\bibfnamefont {J.}~\bibnamefont {Kim}}, \bibinfo {author} {\bibfnamefont {B.~G.}\ \bibnamefont {Park}}, \bibinfo {author} {\bibfnamefont {L.}~\bibnamefont {{\v{S}}mejkal}}, \bibinfo {author} {\bibfnamefont {C.-J.}\ \bibnamefont {Kang}}, \ and\ \bibinfo {author} {\bibfnamefont {C.}~\bibnamefont {Kim}},\ }\href {\doibase 10.1103/PhysRevLett.132.036702} {\bibfield  {journal} {\bibinfo  {journal} {Physical Review Letters}\ }\textbf {\bibinfo {volume} {132}},\ \bibinfo {pages} {036702} (\bibinfo {year} {2024})}\BibitemShut {NoStop}%
\bibitem [{\citenamefont {Krempask{\'{y}}}\ \emph {et~al.}(2024)\citenamefont {Krempask{\'{y}}}, \citenamefont {{\v{S}}mejkal}, \citenamefont {D'Souza}, \citenamefont {Hajlaoui}, \citenamefont {Springholz}, \citenamefont {Uhl{\'{i}}Åov{\'{a}}}, \citenamefont {Alarab}, \citenamefont {Constantinou}, \citenamefont {Strocov}, \citenamefont {Usanov}, \citenamefont {Pudelko}, \citenamefont {Gonz{\'{a}}lez-Hern{\'{a}}ndez}, \citenamefont {{Birk Hellenes}}, \citenamefont {Jansa}, \citenamefont {Reichlov{\'{a}}}, \citenamefont {{\v{S}}ob{\'{a}}Å}, \citenamefont {{Gonzalez Betancourt}}, \citenamefont {Wadley}, \citenamefont {Sinova}, \citenamefont {Kriegner}, \citenamefont {Min{\'{a}}r}, \citenamefont {Dil},\ and\ \citenamefont {Jungwirth}}]{Krempasky2024}%
  \BibitemOpen
  \bibfield  {author} {\bibinfo {author} {\bibfnamefont {J.}~\bibnamefont {Krempask{\'{y}}}}, \bibinfo {author} {\bibfnamefont {L.}~\bibnamefont {{\v{S}}mejkal}}, \bibinfo {author} {\bibfnamefont {S.~W.}\ \bibnamefont {D'Souza}}, \bibinfo {author} {\bibfnamefont {M.}~\bibnamefont {Hajlaoui}}, \bibinfo {author} {\bibfnamefont {G.}~\bibnamefont {Springholz}}, \bibinfo {author} {\bibfnamefont {K.}~\bibnamefont {Uhl{\'{i}}Åov{\'{a}}}}, \bibinfo {author} {\bibfnamefont {F.}~\bibnamefont {Alarab}}, \bibinfo {author} {\bibfnamefont {P.~C.}\ \bibnamefont {Constantinou}}, \bibinfo {author} {\bibfnamefont {V.}~\bibnamefont {Strocov}}, \bibinfo {author} {\bibfnamefont {D.}~\bibnamefont {Usanov}}, \bibinfo {author} {\bibfnamefont {W.~R.}\ \bibnamefont {Pudelko}}, \bibinfo {author} {\bibfnamefont {R.}~\bibnamefont {Gonz{\'{a}}lez-Hern{\'{a}}ndez}}, \bibinfo {author} {\bibfnamefont {A.}~\bibnamefont {{Birk Hellenes}}}, \bibinfo {author} {\bibfnamefont {Z.}~\bibnamefont {Jansa}}, \bibinfo {author} {\bibfnamefont
  {H.}~\bibnamefont {Reichlov{\'{a}}}}, \bibinfo {author} {\bibfnamefont {Z.}~\bibnamefont {{\v{S}}ob{\'{a}}Å}}, \bibinfo {author} {\bibfnamefont {R.~D.}\ \bibnamefont {{Gonzalez Betancourt}}}, \bibinfo {author} {\bibfnamefont {P.}~\bibnamefont {Wadley}}, \bibinfo {author} {\bibfnamefont {J.}~\bibnamefont {Sinova}}, \bibinfo {author} {\bibfnamefont {D.}~\bibnamefont {Kriegner}}, \bibinfo {author} {\bibfnamefont {J.}~\bibnamefont {Min{\'{a}}r}}, \bibinfo {author} {\bibfnamefont {J.~H.}\ \bibnamefont {Dil}}, \ and\ \bibinfo {author} {\bibfnamefont {T.}~\bibnamefont {Jungwirth}},\ }\href {\doibase 10.1038/s41586-023-06907-7} {\bibfield  {journal} {\bibinfo  {journal} {Nature}\ }\textbf {\bibinfo {volume} {626}},\ \bibinfo {pages} {517} (\bibinfo {year} {2024})}\BibitemShut {NoStop}%
\bibitem [{\citenamefont {Osumi}\ \emph {et~al.}(2024)\citenamefont {Osumi}, \citenamefont {Souma}, \citenamefont {Aoyama}, \citenamefont {Yamauchi}, \citenamefont {Honma}, \citenamefont {Nakayama}, \citenamefont {Takahashi}, \citenamefont {Ohgushi},\ and\ \citenamefont {Sato}}]{Osumi2024}%
  \BibitemOpen
  \bibfield  {author} {\bibinfo {author} {\bibfnamefont {T.}~\bibnamefont {Osumi}}, \bibinfo {author} {\bibfnamefont {S.}~\bibnamefont {Souma}}, \bibinfo {author} {\bibfnamefont {T.}~\bibnamefont {Aoyama}}, \bibinfo {author} {\bibfnamefont {K.}~\bibnamefont {Yamauchi}}, \bibinfo {author} {\bibfnamefont {A.}~\bibnamefont {Honma}}, \bibinfo {author} {\bibfnamefont {K.}~\bibnamefont {Nakayama}}, \bibinfo {author} {\bibfnamefont {T.}~\bibnamefont {Takahashi}}, \bibinfo {author} {\bibfnamefont {K.}~\bibnamefont {Ohgushi}}, \ and\ \bibinfo {author} {\bibfnamefont {T.}~\bibnamefont {Sato}},\ }\href {\doibase 10.1103/PhysRevB.109.115102} {\bibfield  {journal} {\bibinfo  {journal} {Physical Review B}\ }\textbf {\bibinfo {volume} {109}},\ \bibinfo {pages} {115102} (\bibinfo {year} {2024})}\BibitemShut {NoStop}%
\bibitem [{\citenamefont {Kriegner}\ \emph {et~al.}(2017)\citenamefont {Kriegner}, \citenamefont {Reichlova}, \citenamefont {Grenzer}, \citenamefont {Schmidt}, \citenamefont {Ressouche}, \citenamefont {Godinho}, \citenamefont {Wagner}, \citenamefont {Martin}, \citenamefont {Shick}, \citenamefont {Volobuev}, \citenamefont {Springholz}, \citenamefont {Hol{\'{y}}}, \citenamefont {Wunderlich}, \citenamefont {Jungwirth},\ and\ \citenamefont {V{\'{y}}born{\'{y}}}}]{Kriegner2017}%
  \BibitemOpen
  \bibfield  {author} {\bibinfo {author} {\bibfnamefont {D.}~\bibnamefont {Kriegner}}, \bibinfo {author} {\bibfnamefont {H.}~\bibnamefont {Reichlova}}, \bibinfo {author} {\bibfnamefont {J.}~\bibnamefont {Grenzer}}, \bibinfo {author} {\bibfnamefont {W.}~\bibnamefont {Schmidt}}, \bibinfo {author} {\bibfnamefont {E.}~\bibnamefont {Ressouche}}, \bibinfo {author} {\bibfnamefont {J.}~\bibnamefont {Godinho}}, \bibinfo {author} {\bibfnamefont {T.}~\bibnamefont {Wagner}}, \bibinfo {author} {\bibfnamefont {S.~Y.}\ \bibnamefont {Martin}}, \bibinfo {author} {\bibfnamefont {A.~B.}\ \bibnamefont {Shick}}, \bibinfo {author} {\bibfnamefont {V.~V.}\ \bibnamefont {Volobuev}}, \bibinfo {author} {\bibfnamefont {G.}~\bibnamefont {Springholz}}, \bibinfo {author} {\bibfnamefont {V.}~\bibnamefont {Hol{\'{y}}}}, \bibinfo {author} {\bibfnamefont {J.}~\bibnamefont {Wunderlich}}, \bibinfo {author} {\bibfnamefont {T.}~\bibnamefont {Jungwirth}}, \ and\ \bibinfo {author} {\bibfnamefont {K.}~\bibnamefont {V{\'{y}}born{\'{y}}}},\ }\href
  {\doibase 10.1103/PhysRevB.96.214418} {\bibfield  {journal} {\bibinfo  {journal} {Physical Review B}\ }\textbf {\bibinfo {volume} {96}},\ \bibinfo {pages} {214418} (\bibinfo {year} {2017})}\BibitemShut {NoStop}%
\bibitem [{\citenamefont {Siol}\ \emph {et~al.}(2018)\citenamefont {Siol}, \citenamefont {Han}, \citenamefont {Mangum}, \citenamefont {Schulz}, \citenamefont {Holder}, \citenamefont {Klein}, \citenamefont {van Hest}, \citenamefont {Gorman},\ and\ \citenamefont {Zakutayev}}]{Siol2018}%
  \BibitemOpen
  \bibfield  {author} {\bibinfo {author} {\bibfnamefont {S.}~\bibnamefont {Siol}}, \bibinfo {author} {\bibfnamefont {Y.}~\bibnamefont {Han}}, \bibinfo {author} {\bibfnamefont {J.}~\bibnamefont {Mangum}}, \bibinfo {author} {\bibfnamefont {P.}~\bibnamefont {Schulz}}, \bibinfo {author} {\bibfnamefont {A.~M.}\ \bibnamefont {Holder}}, \bibinfo {author} {\bibfnamefont {T.~R.}\ \bibnamefont {Klein}}, \bibinfo {author} {\bibfnamefont {M.~F. A.~M.}\ \bibnamefont {van Hest}}, \bibinfo {author} {\bibfnamefont {B.}~\bibnamefont {Gorman}}, \ and\ \bibinfo {author} {\bibfnamefont {A.}~\bibnamefont {Zakutayev}},\ }\href {\doibase 10.1039/C8TC01828F} {\bibfield  {journal} {\bibinfo  {journal} {Journal of Materials Chemistry C}\ }\textbf {\bibinfo {volume} {6}},\ \bibinfo {pages} {6297} (\bibinfo {year} {2018})}\BibitemShut {NoStop}%
\bibitem [{\citenamefont {Mori}\ \emph {et~al.}(2020)\citenamefont {Mori}, \citenamefont {Hatayama}, \citenamefont {Shuang}, \citenamefont {Ando},\ and\ \citenamefont {Sutou}}]{Mori2020}%
  \BibitemOpen
  \bibfield  {author} {\bibinfo {author} {\bibfnamefont {S.}~\bibnamefont {Mori}}, \bibinfo {author} {\bibfnamefont {S.}~\bibnamefont {Hatayama}}, \bibinfo {author} {\bibfnamefont {Y.}~\bibnamefont {Shuang}}, \bibinfo {author} {\bibfnamefont {D.}~\bibnamefont {Ando}}, \ and\ \bibinfo {author} {\bibfnamefont {Y.}~\bibnamefont {Sutou}},\ }\href {\doibase 10.1038/s41467-019-13747-5} {\bibfield  {journal} {\bibinfo  {journal} {Nature Communications}\ }\textbf {\bibinfo {volume} {11}},\ \bibinfo {pages} {85} (\bibinfo {year} {2020})}\BibitemShut {NoStop}%
\bibitem [{\citenamefont {{See Supplemental Material at xx, which includes Refs.[47-49]. We present the first-principles computational details, spin current implement, band structure, and photo-induced spin current and charge current by two different mechanisms of MnTe and BFO with and without SOC}}()}]{supp}%
  \BibitemOpen
  \bibfield  {author} {\bibinfo {author} {\bibnamefont {{See Supplemental Material at xx, which includes Refs.[47-49]. We present the first-principles computational details, spin current implement, band structure, and photo-induced spin current and charge current by two different mechanisms of MnTe and BFO with and without SOC}}},\ }\href@noop {} {\ }\BibitemShut {NoStop}%
\bibitem [{\citenamefont {Kresse}\ and\ \citenamefont {Furthm{\"{u}}ller}(1996)}]{Kresse1996}%
  \BibitemOpen
  \bibfield  {author} {\bibinfo {author} {\bibfnamefont {G.}~\bibnamefont {Kresse}}\ and\ \bibinfo {author} {\bibfnamefont {J.}~\bibnamefont {Furthm{\"{u}}ller}},\ }\href {\doibase 10.1103/PhysRevB.54.11169} {\bibfield  {journal} {\bibinfo  {journal} {Physical Review B}\ }\textbf {\bibinfo {volume} {54}},\ \bibinfo {pages} {11169} (\bibinfo {year} {1996})}\BibitemShut {NoStop}%
\bibitem [{\citenamefont {Dudarev}\ \emph {et~al.}(1998)\citenamefont {Dudarev}, \citenamefont {Botton}, \citenamefont {Savrasov}, \citenamefont {Humphreys},\ and\ \citenamefont {Sutton}}]{Dudarev1998}%
  \BibitemOpen
  \bibfield  {author} {\bibinfo {author} {\bibfnamefont {S.~L.}\ \bibnamefont {Dudarev}}, \bibinfo {author} {\bibfnamefont {G.~A.}\ \bibnamefont {Botton}}, \bibinfo {author} {\bibfnamefont {S.~Y.}\ \bibnamefont {Savrasov}}, \bibinfo {author} {\bibfnamefont {C.~J.}\ \bibnamefont {Humphreys}}, \ and\ \bibinfo {author} {\bibfnamefont {A.~P.}\ \bibnamefont {Sutton}},\ }\href {\doibase 10.1103/PhysRevB.57.1505} {\bibfield  {journal} {\bibinfo  {journal} {Physical Review B}\ }\textbf {\bibinfo {volume} {57}},\ \bibinfo {pages} {1505} (\bibinfo {year} {1998})}\BibitemShut {NoStop}%
\bibitem [{\citenamefont {Perdew}\ \emph {et~al.}(1996)\citenamefont {Perdew}, \citenamefont {Burke},\ and\ \citenamefont {Ernzerhof}}]{perdew1996}%
  \BibitemOpen
  \bibfield  {author} {\bibinfo {author} {\bibfnamefont {J.~P.}\ \bibnamefont {Perdew}}, \bibinfo {author} {\bibfnamefont {K.}~\bibnamefont {Burke}}, \ and\ \bibinfo {author} {\bibfnamefont {M.}~\bibnamefont {Ernzerhof}},\ }\href {\doibase 10.1103/PhysRevLett.77.3865} {\bibfield  {journal} {\bibinfo  {journal} {Phys. Rev. Lett.}\ }\textbf {\bibinfo {volume} {77}},\ \bibinfo {pages} {3865} (\bibinfo {year} {1996})}\BibitemShut {NoStop}%
\bibitem [{\citenamefont {Kiselev}\ \emph {et~al.}(1963)\citenamefont {Kiselev}, \citenamefont {Ozerov},\ and\ \citenamefont {Zhdanov}}]{Kiselev1963DetectionOM}%
  \BibitemOpen
  \bibfield  {author} {\bibinfo {author} {\bibfnamefont {S.~V.}\ \bibnamefont {Kiselev}}, \bibinfo {author} {\bibfnamefont {R.~P.}\ \bibnamefont {Ozerov}}, \ and\ \bibinfo {author} {\bibfnamefont {G.~S.}\ \bibnamefont {Zhdanov}},\ }\href {https://api.semanticscholar.org/CorpusID:137084304} {\bibfield  {journal} {\bibinfo  {journal} {Soviet physics. Doklady}\ }\textbf {\bibinfo {volume} {7}},\ \bibinfo {pages} {742} (\bibinfo {year} {1963})}\BibitemShut {NoStop}%
\bibitem [{\citenamefont {Teague}\ \emph {et~al.}(1970)\citenamefont {Teague}, \citenamefont {Gerson},\ and\ \citenamefont {James}}]{Teague1970}%
  \BibitemOpen
  \bibfield  {author} {\bibinfo {author} {\bibfnamefont {J.~R.}\ \bibnamefont {Teague}}, \bibinfo {author} {\bibfnamefont {R.}~\bibnamefont {Gerson}}, \ and\ \bibinfo {author} {\bibfnamefont {W.}~\bibnamefont {James}},\ }\href {\doibase 10.1016/0038-1098(70)90262-0} {\bibfield  {journal} {\bibinfo  {journal} {Solid State Communications}\ }\textbf {\bibinfo {volume} {8}},\ \bibinfo {pages} {1073} (\bibinfo {year} {1970})}\BibitemShut {NoStop}%
\bibitem [{\citenamefont {Fischer}\ \emph {et~al.}(1980)\citenamefont {Fischer}, \citenamefont {Polomska}, \citenamefont {Sosnowska},\ and\ \citenamefont {Szymanski}}]{Fischer1980a}%
  \BibitemOpen
  \bibfield  {author} {\bibinfo {author} {\bibfnamefont {P.}~\bibnamefont {Fischer}}, \bibinfo {author} {\bibfnamefont {M.}~\bibnamefont {Polomska}}, \bibinfo {author} {\bibfnamefont {I.}~\bibnamefont {Sosnowska}}, \ and\ \bibinfo {author} {\bibfnamefont {M.}~\bibnamefont {Szymanski}},\ }\href {\doibase 10.1088/0022-3719/13/10/012} {\bibfield  {journal} {\bibinfo  {journal} {Journal of Physics C: Solid State Physics}\ }\textbf {\bibinfo {volume} {13}},\ \bibinfo {pages} {1931} (\bibinfo {year} {1980})}\BibitemShut {NoStop}%
\bibitem [{\citenamefont {Young}\ \emph {et~al.}(2012)\citenamefont {Young}, \citenamefont {Zheng},\ and\ \citenamefont {Rappe}}]{Young2-2012}%
  \BibitemOpen
  \bibfield  {author} {\bibinfo {author} {\bibfnamefont {S.~M.}\ \bibnamefont {Young}}, \bibinfo {author} {\bibfnamefont {F.}~\bibnamefont {Zheng}}, \ and\ \bibinfo {author} {\bibfnamefont {A.~M.}\ \bibnamefont {Rappe}},\ }\href {\doibase 10.1103/PhysRevLett.109.236601} {\bibfield  {journal} {\bibinfo  {journal} {Physical Review Letters}\ }\textbf {\bibinfo {volume} {109}},\ \bibinfo {pages} {236601} (\bibinfo {year} {2012})}\BibitemShut {NoStop}%
\bibitem [{\citenamefont {Yang}\ \emph {et~al.}(2010)\citenamefont {Yang}, \citenamefont {Seidel}, \citenamefont {Byrnes}, \citenamefont {Shafer}, \citenamefont {Yang}, \citenamefont {Rossell}, \citenamefont {Yu}, \citenamefont {Chu}, \citenamefont {Scott}, \citenamefont {Ager}, \citenamefont {Martin},\ and\ \citenamefont {Ramesh}}]{Yang2010}%
  \BibitemOpen
  \bibfield  {author} {\bibinfo {author} {\bibfnamefont {S.~Y.}\ \bibnamefont {Yang}}, \bibinfo {author} {\bibfnamefont {J.}~\bibnamefont {Seidel}}, \bibinfo {author} {\bibfnamefont {S.~J.}\ \bibnamefont {Byrnes}}, \bibinfo {author} {\bibfnamefont {P.}~\bibnamefont {Shafer}}, \bibinfo {author} {\bibfnamefont {C.-H.}\ \bibnamefont {Yang}}, \bibinfo {author} {\bibfnamefont {M.~D.}\ \bibnamefont {Rossell}}, \bibinfo {author} {\bibfnamefont {P.}~\bibnamefont {Yu}}, \bibinfo {author} {\bibfnamefont {Y.-H.}\ \bibnamefont {Chu}}, \bibinfo {author} {\bibfnamefont {J.~F.}\ \bibnamefont {Scott}}, \bibinfo {author} {\bibfnamefont {J.~W.}\ \bibnamefont {Ager}}, \bibinfo {author} {\bibfnamefont {L.~W.}\ \bibnamefont {Martin}}, \ and\ \bibinfo {author} {\bibfnamefont {R.}~\bibnamefont {Ramesh}},\ }\href {\doibase 10.1038/nnano.2009.451} {\bibfield  {journal} {\bibinfo  {journal} {Nature Nanotechnology}\ }\textbf {\bibinfo {volume} {5}},\ \bibinfo {pages} {143} (\bibinfo {year} {2010})}\BibitemShut {NoStop}%
\end{thebibliography}%
\end{document}